\documentclass[aps,floats,twocolumn,nofootinbib,preprintnumbers]{revtex4}

\usepackage{amssymb}
\usepackage{amsmath}

\usepackage{graphicx}
\usepackage{graphics}
\usepackage{dcolumn}
\usepackage{color}
\usepackage{tikz}

\usepackage{rotate}
\usepackage{fancyhdr} 
\usepackage{hyperref}
\usepackage{indentfirst}
\usepackage{multirow}
\def\hhref#1{\href{http://arxiv.org/abs/#1}{#1}} 

\usepackage{umoline}
\usepackage{ulem}

\usepackage{hyperref}
\definecolor{oucrimsonred}{rgb}{0.6, 0.0, 0.0}
\definecolor{persianblue}{rgb}{0.11, 0.22, 0.73}
\definecolor{forestgreen}{rgb}{0.13,0.35,0.13}
 \hypersetup{colorlinks, citecolor=oucrimsonred, linkcolor=persianblue, urlcolor=oucrimsonred}

\def\be{\begin{equation}}
\def\ee{\end{equation}}
\def\bea{\begin{eqnarray}}
\def\eea{\end{eqnarray}}
\def\ba{\begin{eqnarray}}
\def\ea{\end{eqnarray}}

\definecolor{darkred}{rgb}{.743,0,0}

\begin{document}

\title{On the fate of the Standard Model at finite temperature}

\author{Luigi Delle Rose\,$^{a,b}$}
\email{luigi.dellerose@le.infn.it}

\author{Carlo Marzo\,$^{a,b}$}
\email{carlo.marzo@le.infn.it}

\author{Alfredo Urbano\,$^{c}$}
\email{alfredo.urbano@sissa.it}

\affiliation{ $^{a}$Universit\`a del Salento, Dipartimento di Matematica e Fisica ``Ennio De Giorgi", Via Arnesano, 73100 Lecce (Italy).\\
$^{b}$INFN - Istituto Nazionale di Fisica Nucleare sez. di Lecce, via Arnesano, 73100 Lecce (Italy).\\
$^{c}$SISSA - International School for Advanced Studies, via Bonomea 256, 34136 Trieste (Italy).}

\preprint{SISSA 34/2015/FISI}

\date{\today}

\begin{abstract}
In this paper we revisit and update the computation of thermal corrections to the stability of the electroweak vacuum in the Standard Model.
At zero temperature, we make use of the full two-loop effective potential, improved by three-loop beta functions with two-loop matching conditions. 
At finite temperature, we include one-loop thermal corrections together with resummation of daisy diagrams. 
We solve numerically---both at zero and finite temperature---the 
bounce equation, thus providing an accurate description of the thermal tunneling. 
Assuming a maximum temperature in the early Universe of the order of $10^{18}$ GeV, we find that the instability bound excludes  
values of the top mass $M_t \gtrsim 173.6$ GeV, with $M_h \simeq 125$ GeV and including uncertainties on the strong coupling.
We discuss the validity and temperature-dependence of this bound in the early Universe, with a special focus on the reheating phase after inflation.
\end{abstract}

\maketitle

\section{Introduction}\label{sec:Intro}

The Standard Model (SM), if extrapolated up to extremely high energies by means of its Renormalization Group (RG) equations, 
reveals a rather peculiar property: the electroweak vacuum does not correspond to the configuration of minimal energy;
contrarily, it is a metastable state close to a phase transition~\cite{Buttazzo:2013uya}.
This scenario---dubbed {\it near-criticality}---must be considered as the most important theoretical message learned from the LHC run I.
Near-criticality may open a window on the realm of Planck-scale physics, otherwise completely inaccessible from a phenomenological point of view.
Understanding its meaning, and refining the computational tools needed to this end, is therefore a task of primary importance.

On the quantitative level, near-criticality emerges from the computation of the tunneling probability---integrated 
over the age of the Universe---between the false and true vacuum of the Higgs potential~\cite{Coleman:1977py,Callan:1977pt}. Probabilities larger than one correspond to an unstable configuration of the electroweak vacuum. In the SM with Higgs mass $M_h \simeq 125$ GeV
the instability occurs if $M_t \gtrsim 178$ GeV, a value of the top mass that is fairly away from present experimental measurements. 
In other words, for the present central values of  $M_h$ and $M_t$,  the electroweak vacuum of the SM is 
unstable but sufficiently long-lived if compared to the age of the Universe. 

However, this result relies on the assumption that thermal effects, due to non-zero values of the temperature, are neglected. 
The impact of thermal corrections on the computation of the tunneling probability  was intensively 
discussed in the past~\cite{Anderson:1990aa,Arnold:1991cv,Espinosa:1995se}. 
Intuitively,  thermal fluctuations at finite temperature increase the tunneling probability, and 
the easiest way to visualize their role is to think about the analogy with the 
one-dimensional quantum mechanical system of a particle in a potential with a false ground state. 
The thermal kinetic energy borrowed from the heat bath shifts the particle from the initial position at the bottom of the false vacuum, 
thus facilitating the tunneling across the potential barrier. In quantum field theory a proper formulation of the problem requires the computation of {\it i)} the finite 
temperature effective potential, 
and {\it ii)} the {\it bounce} field configuration, namely the solution
of the classical equations of motion that triggers the tunneling between the false vacuum and the other side of the potential barrier~\cite{Coleman:1977py,Callan:1977pt}.

Apart from computational technicalities, on the interpretational side thermal corrections may play an important role since it is very likely that  our Universe---in the early stages of its existence---went through an extremely hot phase.
In~\cite{Espinosa:2007qp,Espinosa:2015qea} the instability of the electroweak vacuum was 
investigated from a cosmological perspective (see also~\cite{Kobakhidze:2013tn,Enqvist:2013kaa,Fairbairn:2014zia,Enqvist:2014bua,Kobakhidze:2014xda,Herranen:2014cua,Kamada:2014ufa,Shkerin:2015exa}).
The main emphasis of~\cite{Espinosa:2015qea} was put on the computation of quantum fluctuations of the Higgs field during inflation.
The heart of the matter is that these fluctuations may force the Higgs field to fall down into the true minimum even before inflation ends. 
However, the bottom line is that 
this problematic situation is not realized if the reheating temperature after inflation is sufficiently large. 
As a consequence,~\cite{Espinosa:2015qea} points towards a cosmological scenario 
in which, right after inflation, the Universe is characterized by an extremely high value of the temperature.
Under this condition, thermal corrections to the tunneling probability---as already noticed in~\cite{Espinosa:2015qea}---can not be neglected.

Motivated by this result, in this paper we revisit and update the computation of thermal corrections to the stability of the electroweak vacuum in the SM, and we structure 
our work as follows. In section~\ref{sec:EffV}, we discuss the finite temperature effective potential used in our analysis. 
In section~\ref{sec:Bounce}, closely following the approach of~\cite{Espinosa:1995se}, we compute the bounce solution and the probability of thermal tunneling.
In section~\ref{sec:ThermalPhaseDiagram}, we present our results in terms of the so-called phase diagram of the SM.
 Finally, we conclude in section~\ref{sec:Conclusions}.
In appendix~\ref{app:A}, we provide further detail about the finite temperature effective potential presented in section~\ref{sec:EffV}.

\section{Effective potential at finite temperature}\label{sec:EffV}

As stated in the introduction,
the starting point of our analysis is the effective potential of the SM at finite temperature.
We use the following short-hand notation
\begin{eqnarray}\label{eq:MasterPotential}
V_{\rm eff}(\phi, T) &=& V_0(\phi) + V_{{\rm 1-loop}}(\phi) + V_{{\rm 2-loop}}(\phi) \nonumber \\
&+& V_{{\rm 1-loop}}(\phi,T) + V_{{\rm ring}}(\phi,T)~,
\end{eqnarray}
where the first (second) line refers to $T=0$ ($T\neq 0$), and $\phi$ is the real Higgs field. 
At $T=0$, we include, in addition to the tree level Higgs potential $V_0(\phi)$, one- and two-loop corrections. 
At $T\neq 0$, we include one-loop thermal diagrams and plasma effects, the latter described by one-loop ring resummation of daisy diagrams. 
For completeness, we collect  the explicit expressions in appendix~\ref{app:A}. In appendix~~\ref{app:B} we discuss the validity of the one-loop approximation at finite temperature.
On a more technical level, the effective potential in eq.~(\ref{eq:MasterPotential}) is equipped with the following tools.

\begin{itemize}

\item[$\circ$]  We implement  the RG improvement of the effective potential in eq.~(\ref{eq:MasterPotential}). 
The dimensionless  parameters run according to the three-loop SM RG equations. The running Higgs field is 
\begin{equation}
\phi(t) = e^{\Gamma(t)}\phi~,~~~~\Gamma(t) = \int_{0}^{t} dt^{\prime} \gamma(t^{\prime})~,
\end{equation}
with $\gamma(t)$ the Higgs field anomalous dimension $d\phi(t)/dt = \gamma(t) \phi(t)$.

\item[$\circ$] The matching condition are evaluated at two loops, following~\cite{Buttazzo:2013uya}.

\item[$\circ$] In order to canonically normalize the Higgs kinetic term, 
we introduce the canonical field $\phi_{\rm can} = e^{\Gamma(\phi)}\phi$.\footnote{After renormalization the Higgs field effective lagrangian is 
\begin{equation}
\mathcal{L}_{\rm eff} = \frac{1}{2}e^{2\Gamma(\phi)}(\partial_{\mu}\phi)(\partial^{\mu}\phi) - V_{\rm eff}(e^{\Gamma(\phi)}\phi)~,
\end{equation}
with $\Gamma(\phi)= \int_{M_t}^{\phi}\gamma(\mu)d\ln\mu$.
The canonically normalized Higgs field $\phi_{\rm can}$ is implicitly defined by $d\phi_{\rm can}/d\phi = e^{\Gamma(\phi)}$. 
We use the approximate solution $\phi_{\rm can} \simeq e^{\Gamma(\phi)}\phi$.  
This approximation amounts to take a constant $e^{\Gamma(\phi)}$. It corresponds to $\gamma(\phi) \ll 1$, 
since $d\phi_{\rm can}/d\phi = e^{\Gamma(\phi)}[1+\gamma(\phi)]$. Indeed, we checked that this condition is always verified during the RG evolution.
} 

\item[$\circ$]  Finally, in order to minimize the impact of large logs, the renormalization scale is chosen according to 
\begin{equation}
\mu(t) = \phi_{\rm can}~,
\end{equation}
where the relation with the running parameter $t$ is $\mu(t)=\mu_0 \exp(t)$. 
The scale $\mu_0$ fixes the starting point of the running, and we use as a reference the physical top mass.
From now on, we suppress the subscript $_{\rm can}$.

\end{itemize}
\begin{figure}[!htb!]
\centering
 \includegraphics[width = 0.45 \textwidth]{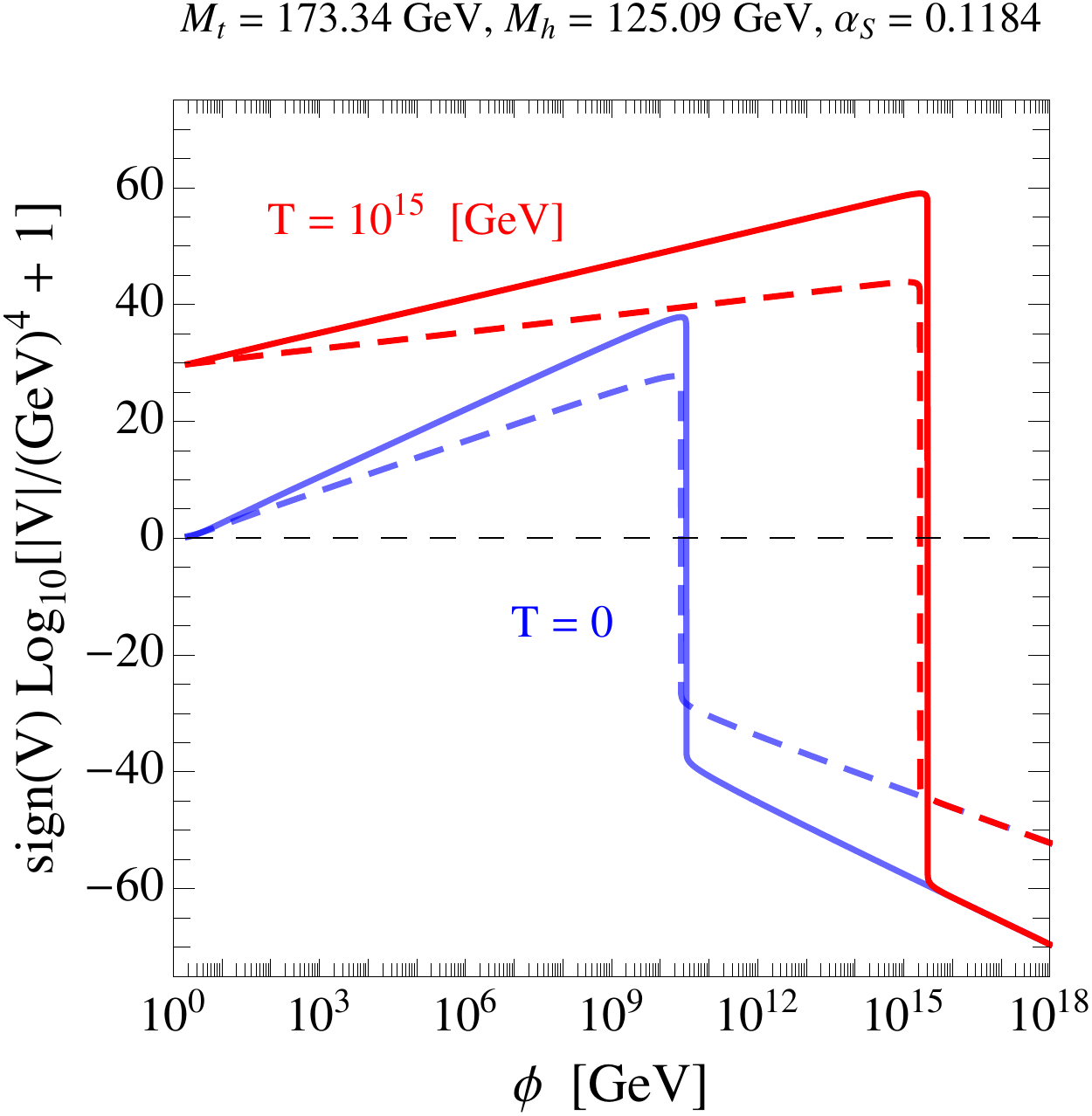}
\caption{ \textit{Effective potential and first derivative at zero and finite temperature as a function of the Higgs field.
Blue line: $T=0$. Red line: finite temperature $T = 10^{15}$ GeV.
Solid line: effective potential. Dashed line: first derivative. The values of the input SM parameters are shown in the plot label.
}}
\label{fig:FullPotential}
\end{figure}
We show the effective potential at zero (blue solid line) and finite (red solid line) temperature in fig.~\ref{fig:FullPotential}.
Dashed lines refer to the corresponding first derivative $dV_{\rm eff}(\phi, T)/d\phi$.
In section~\ref{sec:Bounce}, this derivative enters in the computation of the bounce solution of the euclidean equations of motion.
For the numerical values of the SM input parameters we take $M_W = 80.384$ GeV, $M_Z = 91.1876$ GeV, and $v = 246.22$ GeV 
(respectively, the $W$ and $Z$ pole mass and the vacuum expectation value of the Higgs field).
For the pole Higgs mass we take $M_h = 125.09\pm 0.24$ GeV 
according to the latest combination of both ATLAS and CMS experiments~\cite{Aad:2015zhl}.
Finally, for the strong coupling constant evaluated at $M_Z$ in the $\overline{{\rm MS}}$ scheme (simply $\alpha_s$ hereafter) and the top quark pole mass we take 
$\alpha_s = 0.1184 \pm 0.0007$, $M_t = 173.34 \pm 0.8$. As in~\cite{Buttazzo:2013uya}, the latter is a naive combination of ATLAS, CMS and TeVatron measurements
plagued by unavoidable systematic error due to complicated Monte Carlo modeling~\cite{Alekhin:2012py}. We will come back to this 
point in section~\ref{sec:ThermalPhaseDiagram}.

In fig.~\ref{fig:FullPotential}, notice that at $T = 0$ there is no electroweak minimum since we neglect the quadratic part in the tree level potential (see eq.~(\ref{eq:TreeLevel})).
This approximation is well justified since we are interested in large field values. 
The potential at $T=0$ exhibits the expected behavior, changing sign around field values $\phi \approx 10^{10}$-$10^{11}$ GeV; 
this is the instability scale at which the quartic coupling $\lambda(\phi)$ crosses zero in its RG evolution, and the effective potential develops 
the true vacuum. 
At $T\neq 0$ (for definiteness, we take in fig.~\ref{fig:FullPotential} $T = 10^{15}$ GeV)
thermal corrections dominate over the $T = 0$ part until $\phi \approx T$; for $\phi \gg T$, on the contrary, they are exponentially 
suppressed (see eqs.~(\ref{eq:JB},\ref{eq:JF})), and therefore subdominant if compared with the  $T = 0$ contributions.
\begin{figure}[!htb!]
\centering
 \includegraphics[width = 0.45 \textwidth]{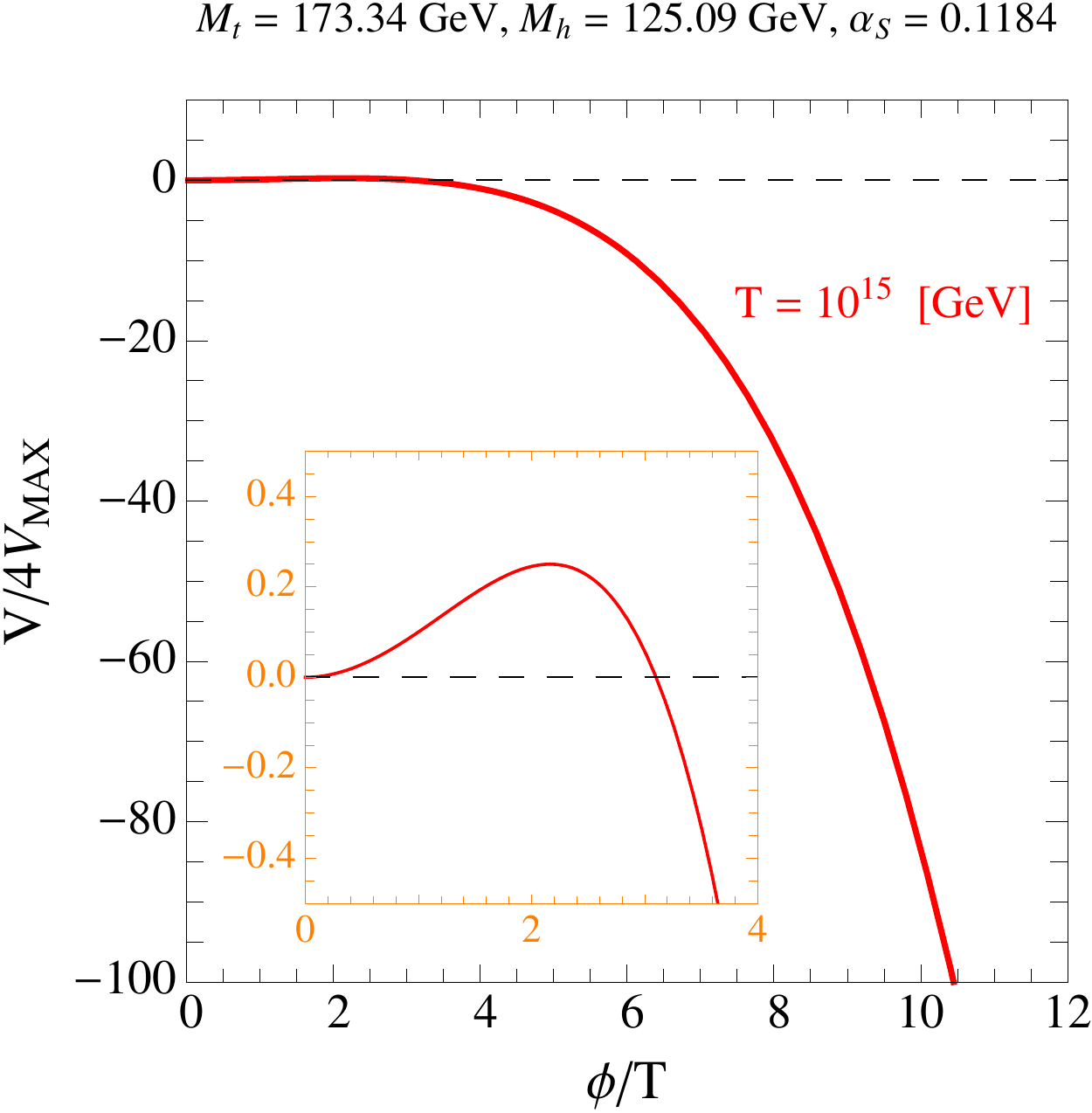}
\caption{ \textit{
SM effective potential (normalized with respect to four times its maximum value) at $T = 10^{15}$ GeV. 
The field $\phi$ scales as $\phi/T$. 
The values of the input SM parameters are shown in the plot label. In the insert, we zoom in the region close to the maximum (axis labels as for the outer plot).
}}
\label{fig:PotPlot}
\end{figure}
The shape of the effective potential at finite temperature can be better visualized in fig.~\ref{fig:PotPlot} where we show the effective potential, normalized with respect to its maximum value, as a function of the Higgs field rescaled according to the ratio $\phi/T$. The effective potential changes sign at about $\phi \simeq 3T$; thenceforth, it sinks towards the true vacuum of the 
theory. Notice that the latter turns out to lie at extremely large field value, $\phi \approx 10^{30}$ GeV~\cite{Branchina:2014rva}. 
However, this is not a problem as soon as one assumes the SM to be valid up to the Planck scale: what 
really matters in terms of tunneling probability---at finite temperature as well as at $T = 0$---is the turning point of the bounce solution rather than the precise location of the true vacuum.
The former, as we shall clarify in the next section, 
never exceeds in our analysis Planck-scale values.

\section{Bounce solution and thermal tunneling}\label{sec:Bounce}

The vacuum decay in a scalar field theory with a potential characterized by an absolute minimum (the true vacuum) and a higher local minimum (the false vacuum)
was first described in~\cite{Coleman:1977py,Callan:1977pt}. The decay proceeds 
via a process called bubble nucleation, that is the tunneling  from a false vacuum field configuration 
to a field configuration---the bounce---containing a bubble of approximate true vacuum embedded in a false vacuum background.

\begin{figure*}[!htb!]
\begin{center}
\centering
  \begin{minipage}{0.45\textwidth}
   \centering
   \includegraphics[scale=0.6]{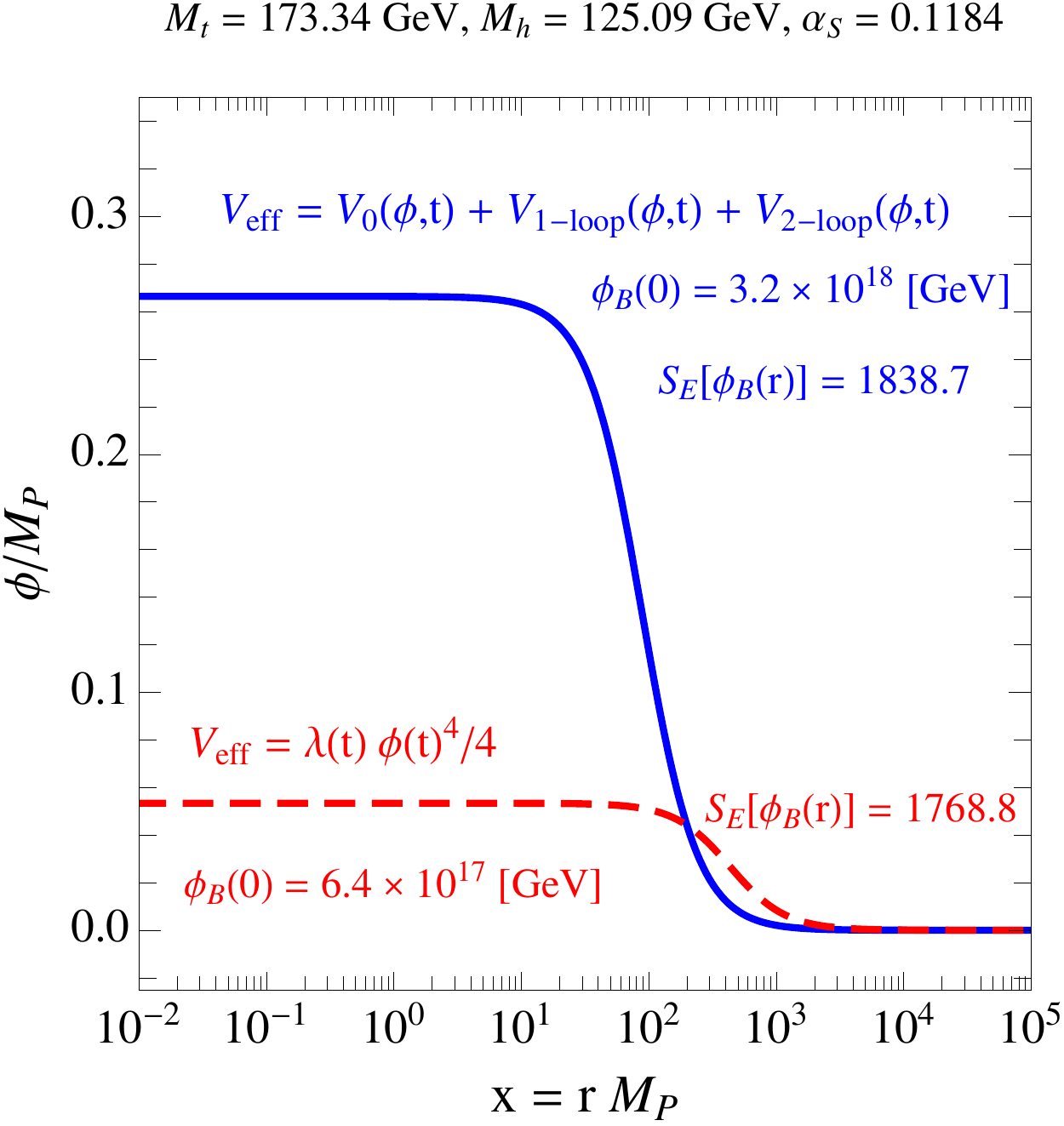}
    \end{minipage}\hspace{0.3 cm}
   \begin{minipage}{0.45\textwidth}
    \centering
    \includegraphics[scale=0.6]{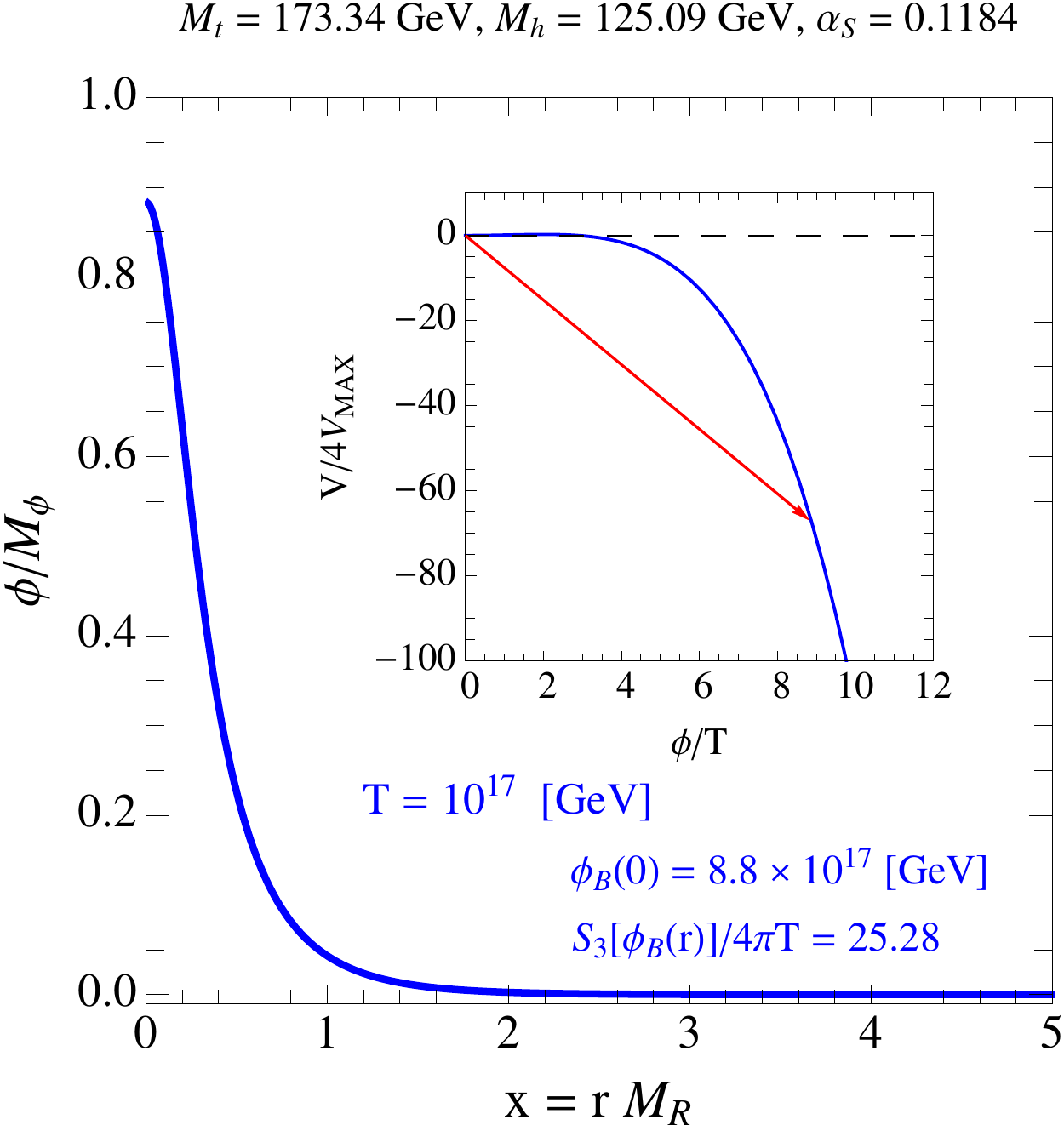}
    \end{minipage}
\caption{\textit{Bounce at $T=0$ (left panel) and $T=10^{17}$ GeV (right panel). 
At zero temperature we show the bounce solution obtained considering a simple tree-level, RG-improved potential (dashed red line) and the full two-loop expression (solid blue line).
The field $\phi$ and the four-dimensional euclidean distance $r$ are rescaled using the Planck mass $M_{\rm P} = 1.22\times 10^{19}$ GeV.
At finite temperature we rescale the field as $\phi(r) = M_{\phi}\times \varphi(r)$, with $M_{\phi} = 10\times T$. 
We rescale the three-dimensional distance according to $x = r\times M_R$, with $M_R \equiv \sqrt{V_{\rm MAX}}/\phi_{\rm MAX}$. 
This prescription greatly improves the efficiency of the numerical shooting method used to solve eq.~(\ref{eq:BounceFiniteT}). 
In the insert, the red arrow pictorially   indicates the bounce solution describing the thermal tunneling. 
The tip of the arrow corresponds to $\phi_B(0) = 8.8\times 10^{17}$ GeV.
The values of the input SM parameters are shown in the plot label.
}}\label{fig:Bounce}
\end{center}
\end{figure*}

At zero temperature the bounce $\phi_B(r)$ is implicitly defined by the differential equation~\cite{Coleman:1977py,Callan:1977pt}
\begin{equation}\label{eq:BounceZeroT}
\frac{d^2\phi}{dr^2} + \frac{3}{r} \frac{d\phi}{dr} = \frac{dV_{\rm eff}(\phi)}{d\phi}~,~~~
\lim_{r \to \infty} \phi(r) = 0~,~~~ \left. \frac{d\phi}{dr}\right|_{r = 0} = 0~,
\end{equation}
with $r^2 \equiv \tau^2 + |\vec{r}|^2$, $\tau$ euclidean time.
It corresponds to a field configuration that sits in the false vacuum at a long euclidean time ago ($\tau \to -\infty$), and 
emerges at rest on the other side of the barrier at time $\tau = 0$~\cite{Coleman:1977py,Callan:1977pt}.
The euclidean action for the $O(4)$ spherically symmetric solution of eq.~(\ref{eq:BounceZeroT}) is
\begin{equation}
S_{\rm E}[\phi(r)] = 2\pi^2\int_0^{\infty}dr\,r^3\left[
\frac{1}{2}\left(
\frac{d\phi}{dr}
\right)^2  + V_{\rm eff}(\phi)
\right]~.
\end{equation}
Let us now focus for the moment on the best-fit values $M_h = 125.09$ GeV, $M_t = 173.34$ GeV, $\alpha_s = 0.1184$.
In the left panel of fig.~\ref{fig:Bounce} we show the SM bounce solution obtained using the tree level RG improved quartic potential (red-dashed line)
and the full two-loop effective potential (blue solid line). 

At finite temperature the bounce solution $\phi_B(r)$ is implicitly defined by~\cite{Anderson:1990aa,Arnold:1991cv,Espinosa:1995se}
\begin{equation}\label{eq:BounceFiniteT}
\frac{d^2\phi}{dr^2} + \frac{2}{r} \frac{d\phi}{dr} = \frac{dV_{\rm eff}(\phi, T)}{d\phi}~,~~~
\lim_{r \to \infty} \phi(r) = 0~,~~~ \left. \frac{d\phi}{dr}\right|_{r = 0} = 0~,
\end{equation}
with now $r \equiv  |\vec{r}|$.
The euclidean action for the $O(3)$ spherically symmetric solution is
\begin{equation}\label{eq:SE}
S_{3}[\phi(r)] = 4\pi \int_0^{\infty}dr\,r^2\left[
\frac{1}{2}\left(
\frac{d\phi}{dr}
\right)^2  + V_{\rm eff}(\phi,T)
\right]~.
\end{equation}
In the right panel of fig.~\ref{fig:Bounce} we show the SM bounce solution  at finite temperature $T = 10^{17}$ GeV. 
Both at zero and finite temperature we solve numerically the bounce equation by means of the shooting method, 
without  any kind of approximation for the effective potential.\footnote{In appendix~\ref{app:A} we compare the full numerical result 
with the approximate solution at finite temperature proposed in~\cite{Arnold:1991cv}, and often used in the literature.} Two comments are in order.

At zero temperature and at the tree level, i.e. considering the potential $V(\phi) = \lambda\phi^4/4$, the bounce solution has the following analytical form
\begin{equation}\label{eq:AnaBounce}
\phi_{B}(r) = \sqrt{\frac{8}{|\lambda|}}\frac{R}{R^2 + r^2}~,
\end{equation}
where $R$ is an arbitrary scale reflecting the scale invariance of the potential. 
This degeneracy is broken by quantum corrections~\cite{Isidori:2001bm}, and only one specific value of $R$---the one saturating the path integral, and defining the size of the bounce, $R_M$ in the following---is singled out.
We can use eq.~(\ref{eq:AnaBounce}) to check the reliability of our  numerical shooting method (dashed red line in the left panel of fig.~\ref{fig:Bounce}). 
First, defining the size of the bounce via $\phi_B(R_M) = \phi_B(0)/2$, we extract $R_M = 434.33\times M_{\rm P}^{-1}$. 
Second, plugging back this number into eq.~(\ref{eq:AnaBounce}), and choosing $\mu = 1/R_M$  for the renormalization 
scale in $\lambda(\mu)$~\cite{Isidori:2001bm,Branchina:2014rva}, 
we indeed find an exact match between our numerical solution and the actual bounce in eq.~(\ref{eq:AnaBounce}).

At finite temperature we rescale---in order to improve 
the efficiency of the numerical shooting algorithm---the field as $\phi(r) = M_{\phi}\times \varphi(r)$, with $M_{\phi} = 10\times T$, and the 
three-dimensional distance according to $x = r\times M_R$, with $M_R \equiv \sqrt{V_{\rm MAX}}/\phi_{\rm MAX}$. 
Throughout our analysis  we always find the relation $\phi_B(0)/T \sim 10$ (see fig.~\ref{fig:Bounce}, right panel, for the specific case with $T = 10^{17}$ GeV).
This is the value of the field configuration at which the bubble of true vacuum is nucleated. 
The red arrow in the insert plot in the right panel of fig.~\ref{fig:Bounce} pictorially represents the bounce solution, with the tip at $\phi_B(0)$.

\begin{figure*}[!htb!]
\begin{center}
\centering
  \begin{minipage}{0.45\textwidth}
   \centering
   \includegraphics[scale=0.6]{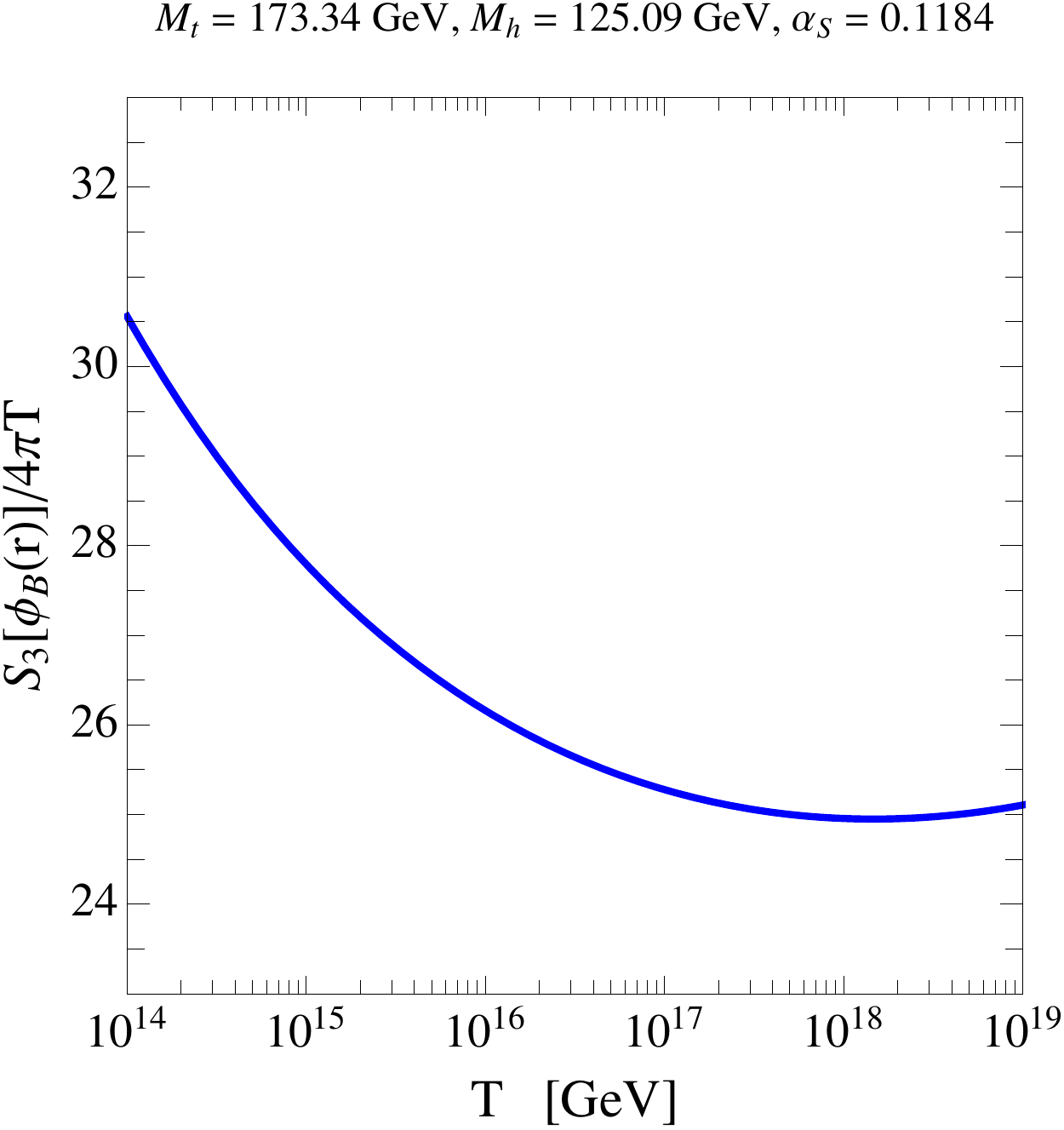}
    \end{minipage}\hspace{0.3 cm}
   \begin{minipage}{0.45\textwidth}
    \centering
    \includegraphics[scale=0.6]{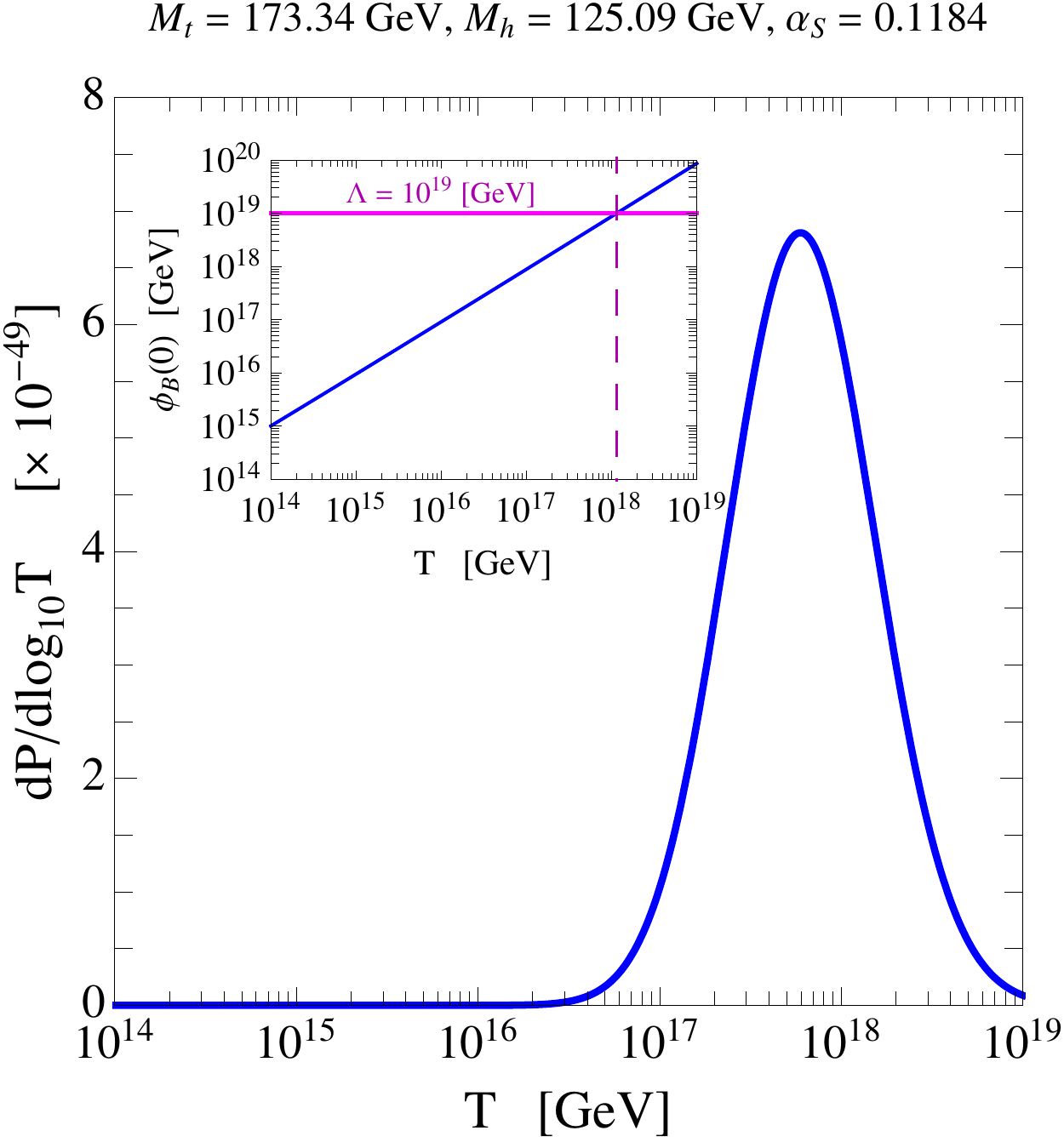}
    \end{minipage}
\caption{\textit{
Left panel. Euclidean action of the bounce solution $\phi_B(r)$ as a function of the temperature.
Right panel. Plot of the differential probability $dP/d\log_{10}T$ as a function of the temperature. 
In the insert, we show the value of $\phi_B(0)$ as a function of the temperature.
For a given cut-off scale (for instance, $\Lambda = 10^{19}$ GeV, solid horizontal magenta line)
the integration of $dP/d\log_{10}T$ must be cut-offed 
at the temperature satisfying the condition $\phi_B(0)\sim \Lambda$ (in this example $T_{\rm cut-off}\sim 10^{18}$ GeV, vertical dashed magenta line).
The values of the input SM parameters are shown in the plot label.
}}
\label{fig:BounceEnergy}
\end{center}
\end{figure*}

The vacuum decay rate per unit volume at fixed temperature $T$ is~\cite{Anderson:1990aa,Arnold:1991cv,Espinosa:1995se}
\begin{equation}
\Gamma(T) \simeq T^4\left\{
\frac{S_{3}[\phi_B(r)]}{2\pi T}
\right\}^{3/2}e^{-S_{3}[\phi_B(r)]/T}~,
\end{equation}
where $E_B \equiv S_{3}[\phi_B(r)]$ represents the energy of a bubble of critical size.
In the left panel of fig.~\ref{fig:BounceEnergy} we show the euclidean action of the bounce solution $\phi_B(r)$ 
as a function of the temperature for the best-fit values of $M_h$, $M_t$, and $\alpha_s$.
The differential decay probability of nucleating a bubble at a given temperature $T$ is given by~\cite{Espinosa:2007qp}
\begin{equation}\label{eq:ProbRadiation}
\frac{dP}{d\ln T} \simeq \Gamma(T) \frac{M_{\rm P}}{T^2}  \left(\frac{\tau_{\rm U}T_0}{T}\right)^3~,
\end{equation}
with $T_0 \simeq 2.35\times 10^{-4}$ eV and $\tau_{\rm U}$ the age of the Universe.
Notice that this formula is valid only in a radiation-dominated Universe.
In the right panel of fig.~\ref{fig:BounceEnergy} 
we show the differential probability $dP/d\log_{10}T$ as a function of the temperature.
 The total integrated probability is defined as
\begin{equation}\label{eq:ThermalProb}
P(T_{\rm cut-off}) = \int_0^{T_{\rm cut-off}} \frac{dP(T^{\prime})}{dT^{\prime}}dT^{\prime}~.
\end{equation}
$T_{\rm cut-off}$ is the cut-off temperature obtained imposing the condition $\phi_B(0) = \Lambda$, where $\Lambda$ 
is the cut-off scale of the SM, for the moment assumed to be $\Lambda= 10^{19}$ GeV.  
In the insert plot in the right panel of fig.~\ref{fig:BounceEnergy} we show the values of $\phi_B(0)$ at different temperatures.
The cut-off at $\Lambda= 10^{19}$ GeV corresponds to a maximum cut-off value on the temperature $T_{\rm cut-off} \simeq 10^{18}$ GeV, as expected since $\phi_B(0)/T \sim 10$.
Larger values of $\phi_B(0) = \Lambda$ would correspond to 
a Planck-scale dominated tunneling transition.
The cut-off temperature plays a fundamental role in connection with the thermal history of the Universe. 
In section~\ref{sec:Reheating} we will discuss this aspect in detail. For the moment, in order to keep the discussion as simple as possible,
we stick to the value  $\Lambda= 10^{19}$ GeV.
Integrating the differential probability using eq.~(\ref{eq:ThermalProb}), we find $P(T_{\rm cut-off}) = 5.22\times10^{-49}\ll 1$.
Consequently, we conclude that the electroweak vacuum of the SM 
for the present central values of  $M_h$, $M_t$, and $\alpha_s$ is 
unstable but sufficiently long-lived if compared to the age of the Universe, even including thermal corrections with the highest cut-off scale $\Lambda= 10^{19}$ GeV.\footnote{Notice that 
the probability defined in eq.~(\ref{eq:ThermalProb}) is not normalized to one, and therefore---strictly speaking---it can not be 
interpreted as a probability in the usual sense.
The correct interpretation was given in~\cite{Guth:1981uk} where it was shown that 
in the bubble nucleation process 
the fraction of space in the false metastable vacuum configuration
 is given by $f_{\rm false} = e^{-P}$, while the fraction of space in the configuration of true vacuum is given by $f_{\rm true} = 1 - e^{-P}$.
 Therefore if $P \ll 1$ ($P \gg 1$) all the space is in the false (true) vacuum field configuration.}

The total probability computed before turns out to be much larger than the corresponding probability evaluated at $T=0$, that is $\sim 10^{-500}$~\cite{Branchina:2014rva}.
Said differently, the electroweak vacuum is still metastable but thermal corrections greatly enhance the tunneling probability.
It simply implies that---extending the previous computation 
to different values of $M_h$, $M_t$, and $\alpha_s$ in the allowed experimental ranges---the resulting instability bound will be much more stringent if compared with the one obtained at $T=0$.
A comprehensive analysis in the context of the phase diagram of the SM will be carried out in section~\ref{sec:ThermalPhaseDiagram}.
For the moment, as a warm-up discussion, let us now try to change only the value of $M_t$.
\begin{figure}[!htb!]
\centering
 \includegraphics[width = 0.45 \textwidth]{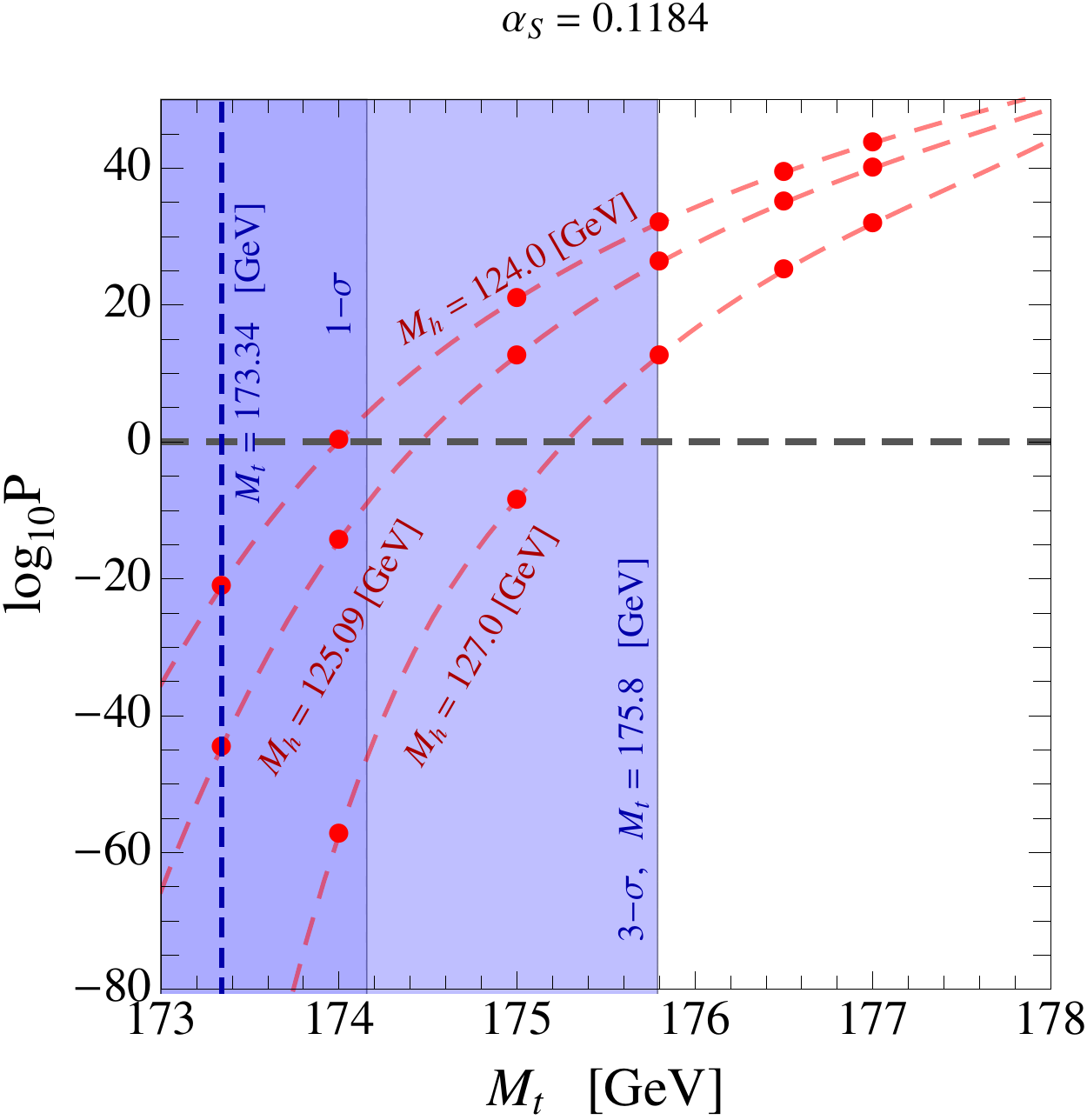}
\caption{ \textit{
Plot of the $\log_{10}$ of the total probability as a function of the top mass for three different values of the Higgs mass, $M_h = 124.0, 125.09, 127.0$ GeV. 
For illustrative purposes, the vertical blue lines mark the best-fit, $1$- and $3$-$\sigma$ 
values according to $M_t = 173.34\pm 0.8$ GeV.
}}
\label{fig:TopProbability}
\end{figure}
In fig.~\ref{fig:TopProbability} we show how the total probability of thermal tunneling changes as a function of 
$M_t$ for three different values of the Higgs mass, $M_h = 124.0, 125.09, 127.0$ GeV, with $\alpha_s = 0.1184$. 
The total probability increases going towards larger values of $M_t$, and smaller values of $M_h$. 
For illustrative purposes, we show the region corresponding to the best-fit, $1$- and $3$-$\sigma$ 
confidence regions of $M_t$ according to $M_t = 173.34\pm 0.8$ GeV. 
For $M_h = 125.09$ GeV, we find that the total probability of thermal tunneling equals one for values of $M_t$ extremely close to the $1$-$\sigma$ confidence region.
This is a remarkable result, given that at $T=0$ the instability bound is reached only for $M_t \gtrsim 178$ GeV. 
Motivated by this result, we turn attention to the full phase diagram of the SM.

\section{The phase diagram of the Standard Model at finite temperature}\label{sec:ThermalPhaseDiagram}

The phase diagram of the SM 
is divided in three regions describing absolute stability, metastability, and instability of the electroweak vacuum 
depending on the values of the SM parameters. 
Among them, the top mass, the Higgs mass, and the strong coupling at weak scale play a dominate role. 
At finite temperature, we add a fourth region in order to discriminate between 
instability at $T=0$ and thermal instability.
All in all, the four regions are defined as follows.

\begin{itemize}

\item[$\circ$] The absolute stability region (green) verifies the condition $\lambda_{\rm eff}(\phi) \geqslant 0$ all the way up to the Planck scale.\footnote{$\lambda_{\rm eff}$ is the effective quartic coupling accounting for one- and two-loop corrections which is extracted from the RG-improved effective potential.} 
The effective potential does not develop a second, deeper minimum, and the electroweak vacuum is stable. 

\item[$\circ$] The instability region at finite temperature (red) 
verifies the condition $P\geqslant 1$, where the thermal tunneling probability is given in eq.~(\ref{eq:ThermalProb}).

\item[$\circ$] At $T=0$, the instability region (marked by the dashed red line) corresponds to a zero-temperature  tunneling 
probability 
\begin{equation}
p = {\rm max}_{R}\frac{V_{\rm U}}{R^4}\exp\left[
-\frac{8\pi^2}{3|\lambda(\mu)|}
\right] > 1~,
\end{equation}
 where $\tau_{\rm U}$ is the age of the Universe and $V_{\rm U} \sim \tau_{\rm U}^4$.

\item[$\circ$]  In the metastability region (yellow) $\lambda_{\rm eff}(\phi)$ does become negative below the Planck scale, and 
the effective potential develops a second minimum deeper than the electroweak one. 
However, the decay probability verifies $P < 1$.

\end{itemize} 

In section~\ref{sec:FullT}---as a natural continuation of what already discussed in section~\ref{sec:Bounce}---we show the phase diagram of the SM at finite temperature 
with the highest cut-off $\Lambda = 10^{19}$ GeV, corresponding to $T_{\rm cut-off} \simeq 10^{18}$ GeV.
In section~\ref{sec:Reheating} we discuss the role of $T_{\rm cut-off}$ in the early Universe, 
thus assessing under which conditions the instability bound at finite temperature applies.

\subsection{Instability bound at finite temperature}\label{sec:FullT}

\begin{figure}[!htb!]
\centering
 \includegraphics[width = 0.45 \textwidth]{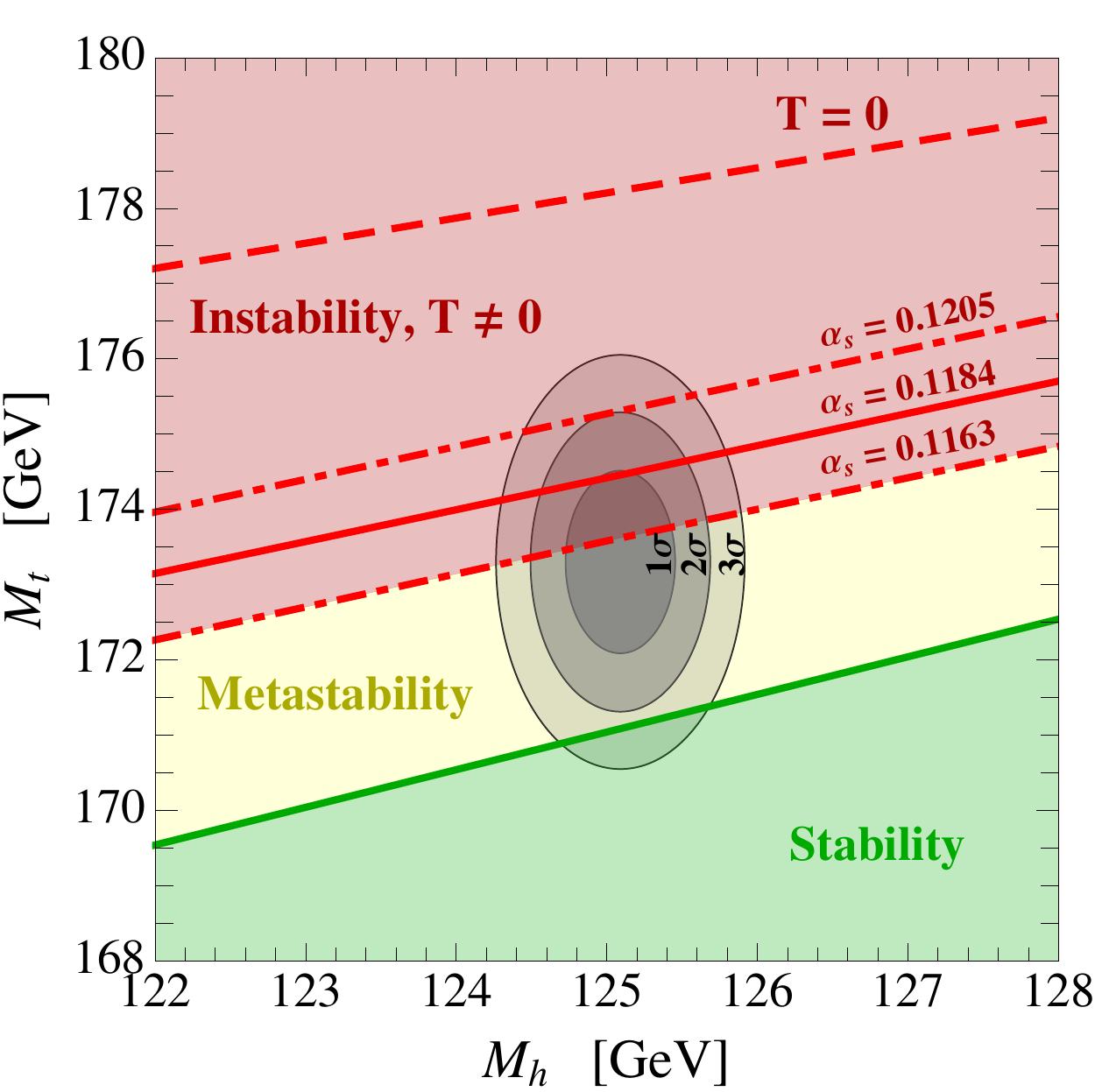}
\caption{ \textit{
SM phase diagram at finite temperature and cut-off scale $\Lambda = 10^{19}$ GeV. Solid (dashed) red line: instability bound with (without) thermal corrections. We also show the $1$-, $2$-, and $3$-$\sigma$ ellipses corresponding to
$M_t = 173.3\pm 0.8$ GeV and $M_h = 125.09\pm 0.24$ GeV (assuming a two-dimensional gaussian distribution without correlations).}}
\label{fig:ThermalPhaseDiagram}
\end{figure}

In fig.~\ref{fig:ThermalPhaseDiagram} we show the phase diagram of the SM in terms of the Higgs and top mass.
The gray ellipses refer to the $1$-, $2$-, and $3$-$\sigma$ confidence regions obtained considering $M_t = 173.3\pm 0.8$ GeV and $M_h = 125.09\pm 0.24$ GeV.
At $T=0$, the instability bound correctly reproduce the known result~\cite{Buttazzo:2013uya} 
according to which, for instance, values $M_t \gtrsim 178$ GeV are excluded if $M_h \simeq 125$ GeV.

At finite temperature, the situation drastically changes. As expected, the instability bound is pushed towards lower values of $M_t$.
To fix the ideas, values $M_t \gtrsim 174.5$ GeV are excluded if $M_h \simeq 125$ GeV. 
Including the uncertainties on the strong coupling at the weak scale (dot-dashed lines in fig.~\ref{fig:ThermalPhaseDiagram}) the
bound becomes even more stringent, and values $M_t \gtrsim 173.6$ GeV are excluded if $M_h \simeq 125$ GeV and $\alpha_s = 0.1163$.

\begin{figure}[!htb!]
\centering
 \includegraphics[width = 0.45 \textwidth]{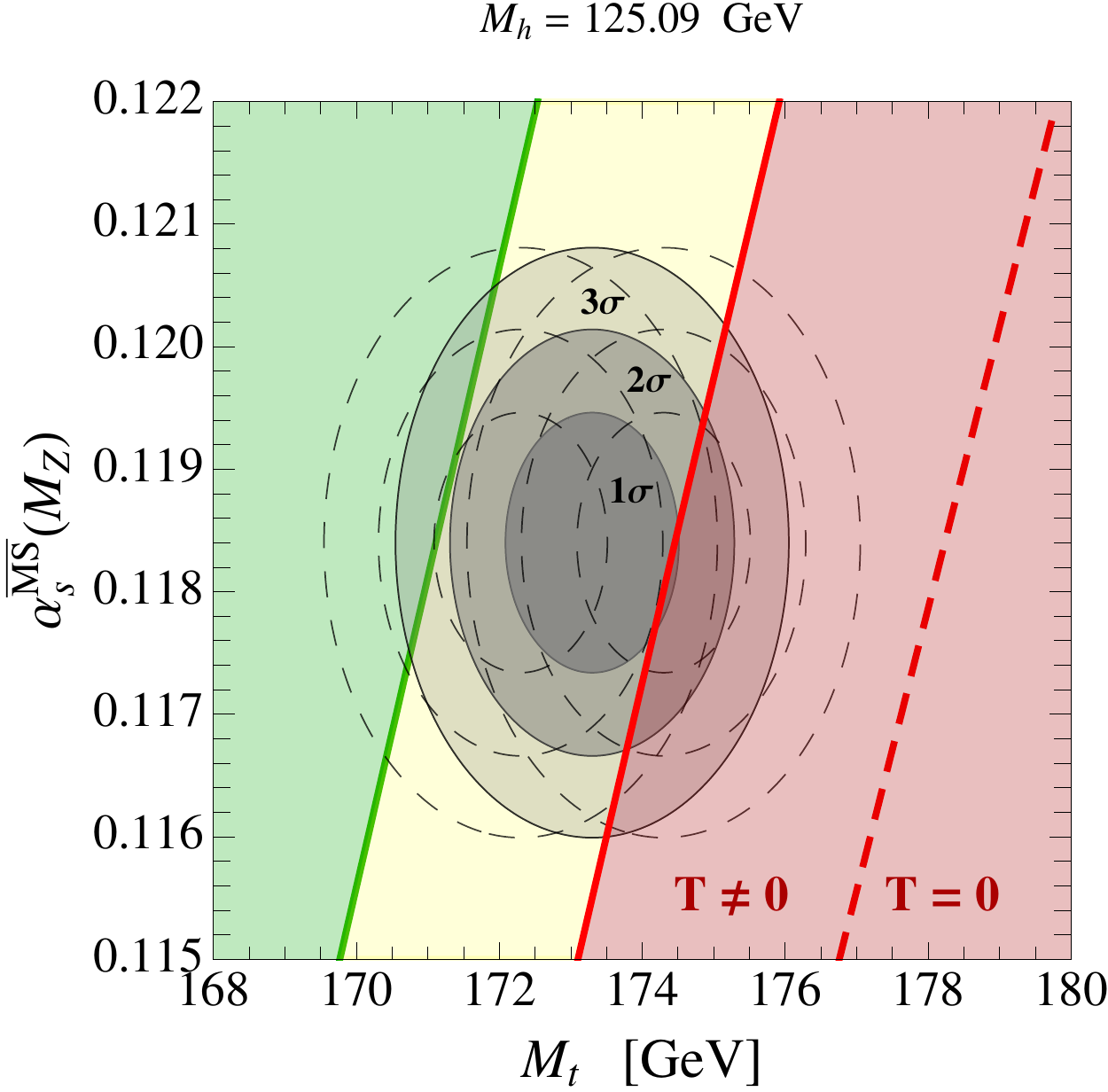}
\caption{ \textit{
Same as in fig.~\ref{fig:ThermalPhaseDiagram} but in the plane $(M_t, \alpha_s)$.
We also show the effect of a $1$ GeV shift in the determination of the top pole mass (dashed ellipses). The Higgs mass is fixed at $M_h = 125.09$ GeV.
}}
\label{fig:ThermalPhaseDiagramAlfa}
\end{figure}

At finite temperature, and assuming the highest cut-off scale $\Lambda = 10^{19}$ GeV, the instability bound 
excludes, taking into account the present experimental uncertainties on $\alpha_s$, almost one half of the 
allowed experimental range for $(M_h, M_t)$. 
In terms of $M_t$ we extract the following bound
 \begin{equation}\label{eq:BoundMaxT}
       \boxed{
       \begin{aligned}
\frac{M_t}{{\rm GeV}} &< 174.459 + 0.4285\times \left(\frac{M_h}{{\rm GeV}} - 125.09\right)   \\
&+ 0.283 \times \left(
\frac{\alpha_s - 0.1184}{0.0007}
\right)
       \end{aligned}
       }
    \end{equation}
In fig.~\ref{fig:ThermalPhaseDiagramAlfa} we show the phase diagram of the SM in terms of the top mass and the strong coupling at the weak scale, keeping $M_h$ fixed at $M_h = 125.09$ GeV.
As before, the ellipses mark the $1$-, $2$- and $3$-$\sigma$ confidence regions with $M_t$ as in fig.~\ref{fig:ThermalPhaseDiagram} and $\alpha_s = 0.1184\pm 0.0007$.
For illustrative purposes, we also show (dashed ellipses) the effect of a $1$ GeV shift in the determination of the top pole mass.
Such shift  symbolically  represents the systematic error involved in the naive combination 
of ATLAS, CMS and TeVatron results used in this paper, $M_t = 173.34 \pm 0.8$ GeV.
Moreover, one should always keep in mind that the experimentally measured top mass is not the pole mass entering in the computation of the 
instability bound but the outcome of a complicated reconstruction of top quark decays (often dubbed the Monte Carlo mass). This fact amounts to a further source of uncertainty.
As well known, and emphasized in this plot, 
the measurement of the top quark pole mass plays a crucial role in the determination of the actual position of the SM in the phase diagram~\cite{Bezrukov:2014ina}.
With the inclusion of thermal corrections, the situation becomes even more severe if compared with the $T= 0$ case, 
since now a small shift of the measured values can drastically change the phase of the electroweak vacuum in both directions, towards the stability as well as the instability region.
This result motivates the need of a future high-energy electron-positron collider, where $M_t$ could be unambiguously measured with a few hundred MeV accuracy in the scattering process $e^+e^- \to t\bar{t}$~\cite{Alekhin:2012py}.

\subsection{Instability bound and reheating temperature}\label{sec:Reheating}

Thermal corrections are computed assuming the Higgs field in equilibrium 
with a thermal bath at temperature $T$.
The occurrence of this condition strongly depends on the 
thermal history of the Universe. During inflation~\cite{Linde:2005ht} all the energy is stored in the inflaton field, which slowly rolls down towards the minimum of its effective potential.
Once reached, inflation ends, and the  inflaton  begins  to
oscillate near the minimum. SM particles are created because of 
their  interactions  with  the inflaton  field: the  kinetic energy  of the oscillating inflaton is gradually transferred into the ultra-relativistic SM particles produced in the final state of its decay.
Eventually, SM particles reach a state of thermal equilibrium at the temperature $T_{\rm RH}$, dubbed reheating temperature~\cite{Allahverdi:2010xz}. 
Thenceforth, the temperature scales according to $T\propto a^{-1}$, as in the ordinary radiation-dominated phase (as customary, $a$ is the Friedmann-Robertson-Walker scale factor).
Strictly speaking, the applicability of our computation is limited to $T < T_{\rm RH}$. In order to further investigate this important point, we explore two possible scenarios.

\subsubsection{Instantaneous reheating}

\begin{figure}[!htb!]
\centering
 \includegraphics[width = 0.45 \textwidth]{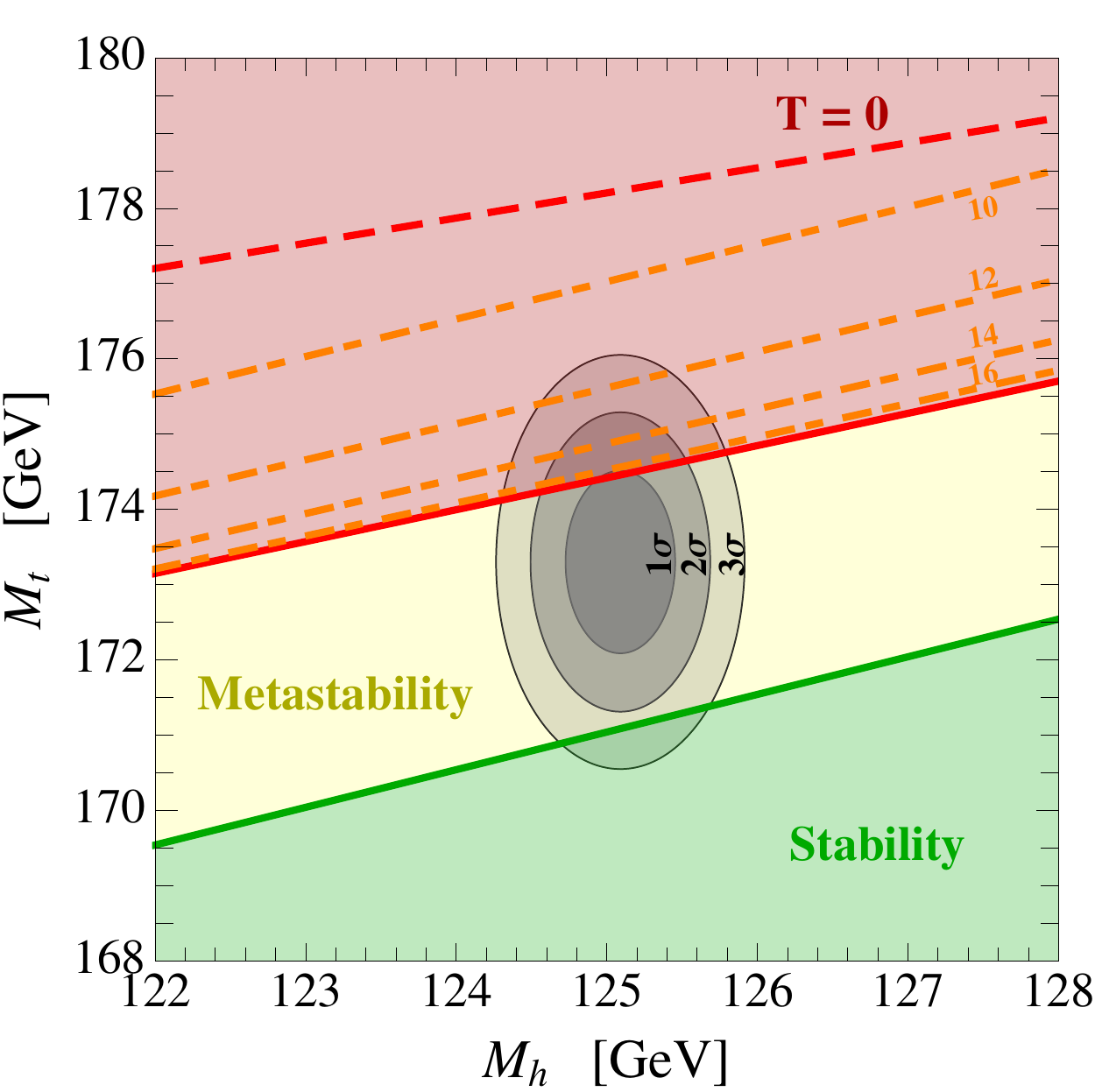}
\caption{ \textit{
Dependence of the instability bound on the reheating temperature $T_{\rm RH}$, assuming instantaneous reheating. 
The orange dot-dashed lines
correspond to different values $T_{\rm RH} = 10^x$ GeV, with---from top to bottom---$x=10,12,14,16$.}}
\label{fig:ThermalPhaseDiagramBis}
\end{figure}

We start describing the reheating as an instantaneous process. 
In this case the decay probability is given by  eq.~(\ref{eq:ThermalProb}), with $T_{\rm cut-off} = T_{\rm RH}$.
In fig.~\ref{fig:ThermalPhaseDiagramBis} we show how the instability bound changes for different values of $T_{\rm RH}$. 
As clear from the right panel of fig.~\ref{fig:BounceEnergy}, 
the largest contribution to the total probability comes from the high-temperature region, and 
a decrease in the cut-off quickly weakens the instability bound.
We show the impact of different reheating temperatures in fig.~\ref{fig:ThermalPhaseDiagramBis}. 
At $T_{\rm RH} \simeq 10^{12}$ GeV 
the instability bound is pushed towards the border of the $3$-$\sigma$ band on $(M_h,M_t)$. 
For smaller values of the reheating temperature, e.g. $T_{\rm RH} = 10^{10}$ GeV, the SM reenters in the metastability region.
The bounds in fig.~\ref{fig:ThermalPhaseDiagram} are well described by the following parametric formula
 \begin{equation}\label{eq:BoundTRH}
       \boxed{
       \begin{aligned}
&\frac{M_t}{{\rm GeV}} < 0.283\times \left(
\frac{\alpha_s- 0.1184}{0.0007}
\right)  \\
&+ c_1\times \frac{M_h}{{\rm GeV}} + c_2\times \log_{10}\frac{T_{\rm RH}}{{\rm GeV}}  + \frac{c_3}{c_4\times \log_{10}\frac{T_{\rm RH}}{{\rm GeV}} + c_5} 
       \end{aligned}
       }
    \end{equation}
with $c_1 = 0.4612$, $c_2 = 1.907$, $c_3 = -1.2\times 10^3$, $c_4 = -0.323$, $c_5 = -8.738$. 
In concrete, taking $M_h = 125.09$ GeV, $\alpha_s = 0.1163$ (close to the $3$-$\sigma$ lower bound), 
and $T_{\rm RH} = 10^{16}$ GeV we find 
$M_t < 173.65$ GeV.

Before proceeding, let us pause for a moment to comment about the current experimental limits on the reheating temperature. 
Despite its relevance in our understanding of the early Universe, very little is known about the actual value of the reheating temperature.
An obvious  lower bound can be obtained 
requiring a successful Big Bang Nucleosynthesis, and it turns out to be $T_{\rm RH} \gtrsim 10$ MeV~\cite{Abbott:1986kb}.
As far as the upper bound is concerned, it is possible---assuming instantaneous reheating---to relate 
the reheating temperature to the energy scale of the inflationary potential~\cite{Ade:2015lrj};
since the latter can be constrained using the limit on the tensor-to-scalar ratio of the amplitudes produced during inflation,
it is possible to extract a bound on $T_{\rm RH}$. All in all, one finds $T_{\rm RH}\lesssim 10^{16}$ GeV~\cite{Ade:2015lrj}.
High values of reheating temperature---as large as the ones considered in fig.~\ref{fig:ThermalPhaseDiagramBis}---are therefore experimentally allowed.
Moreover, the hypothesis of instantaneous reheating is a crude, yet not unrealistic, approximation. 
More likely, reheating is a dynamical process. In the next section we will elaborate this point and its consequences in more detail.

\subsubsection{Including the dynamics of reheating}

\begin{figure}[!htb!]
\centering
 \includegraphics[width = 0.45 \textwidth]{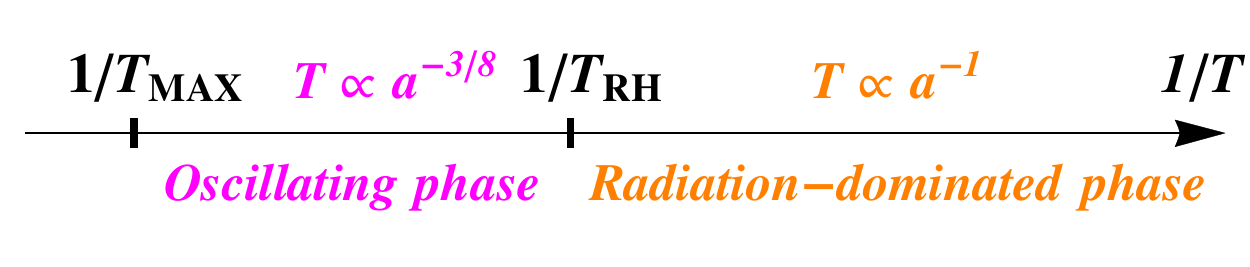}
\caption{ \textit{
Schematic representation of the thermal evolution of the Universe after inflation.
At the end of the reheating process ($T < T_{\rm RH}$) the temperature scales according to $T\propto a^{-1}$, as in the ordinary radiation-dominated phase.
During the oscillating phase of the inflaton, before reheating is completed, $T\propto a^{-3/8}$.
}}
\label{fig:Arrow}
\end{figure}

\begin{figure}[!htb!]
\centering
 \includegraphics[width = 0.45 \textwidth]{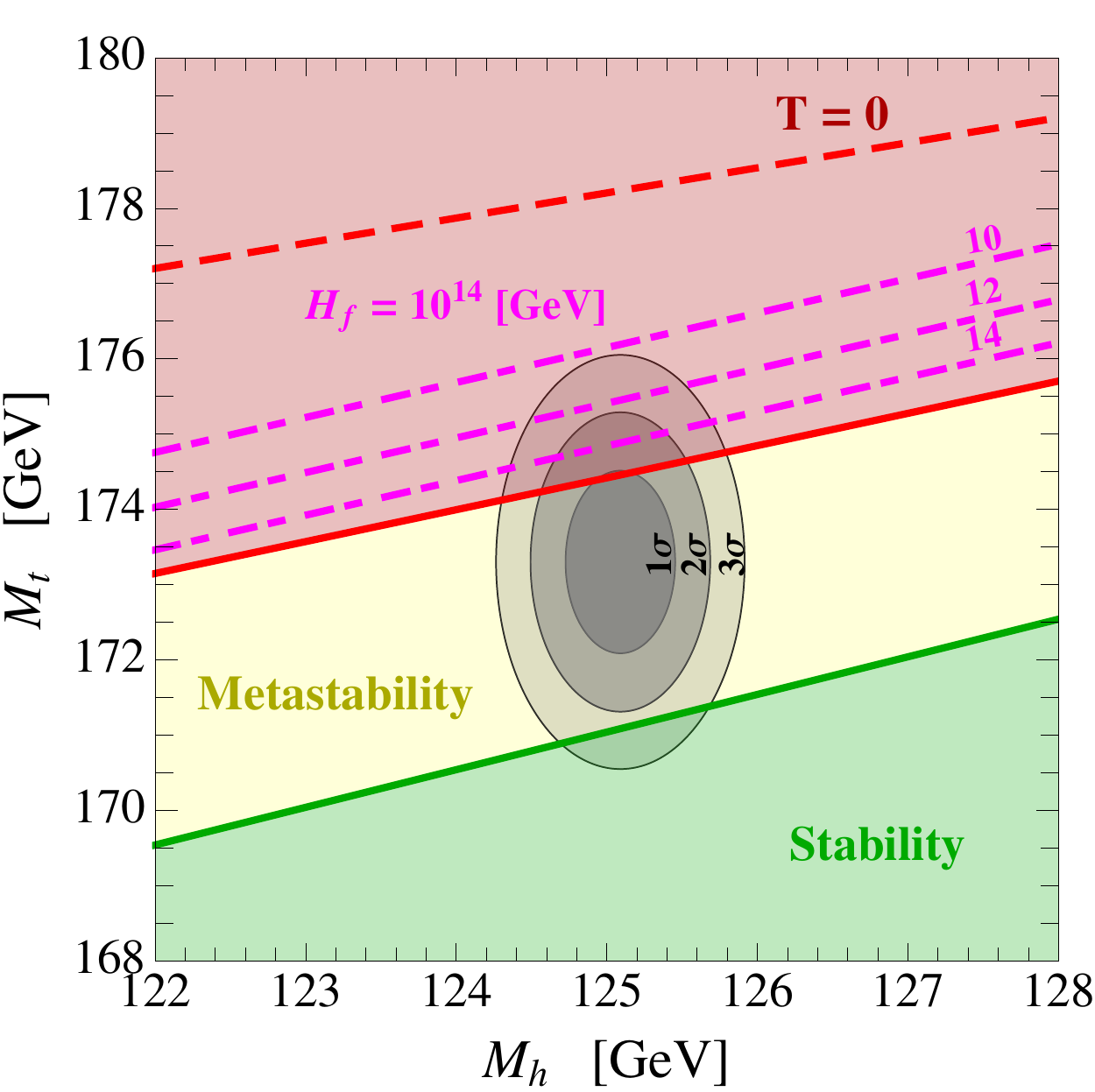}
\caption{ \textit{
Dependence of the instability bound on the reheating temperature $T_{\rm RH}$, including the dynamics of reheating. 
The magenta dot-dashed lines
correspond to different values $T_{\rm RH} = 10^x$ GeV, with---from top to bottom---$x=10,12,14$. We take $H_f = 10^{14}$ GeV and $g_*(T) = 106.75$.}}
\label{fig:OscillatingPhaseDiagram}
\end{figure}

Reheating is not an instantaneous process. On the contrary, the radiation-dominated phase at $T < T_{\rm RH}$ follows a stage of matter domination 
during which the energy density of the Universe is
dominated by the oscillations of the inflaton field~\cite{Scherrer:1984fd,Giudice:2000ex}. Temperature scales according to $T\propto a^{-3/8}$~\cite{Scherrer:1984fd,Giudice:2000ex}; in other words, 
during the oscillating phase the Universe cools down more slowly---if compared with the scaling $T\propto a^{-1}$ of the radiation-dominated phase---because of the heating effect of the inflaton decay. 
As shown in~\cite{Scherrer:1984fd,Giudice:2000ex} the maximum value of the temperature is
\begin{eqnarray}\label{eq:TMAX}
&& T_{\rm MAX} =\\ && \left(\frac{3}{8}\right)^{2/5}\left(\frac{5}{\pi^3}\right)^{1/8}\frac{g_*^{1/8}(T_{\rm RH})}{g_*^{1/4}(T_{\rm MAX})}(M_{\rm P}H_fT_{\rm RH}^{2})^{1/4}~,\nonumber
\end{eqnarray}
where $g_*(T)$ is the effective number of degrees of freedom, and $H_f$ is the Hubble parameter at the end of inflation. 
The situation is schematically summarized in fig.~\ref{fig:Arrow}. 
In the region $T_{\rm RH} \leqslant T \leqslant T_{\rm MAX}$ we can not compute the decay
 probability using eq.~(\ref{eq:ProbRadiation}),
since it relies on the assumption of a radiation-dominated Universe. Using the scaling $T\propto a^{-3/8}$, in the region $T_{\rm RH} \leqslant T \leqslant T_{\rm MAX}$
the differential decay probability becomes~\cite{Espinosa:2007qp}
\begin{equation}\label{eq:ProbOscillation}
\frac{dP}{d\ln T} \simeq \Gamma(T) \frac{M_{\rm P}}{T^2}  \left(\frac{\tau_{\rm U}T_0}{T_{\rm RH}}\right)^3\left(\frac{T_{\rm RH}}{T}\right)^{10}~.
\end{equation}
All in all, the total integrated probability is given by
\begin{eqnarray}
P(T_{\rm RH}, H_f) &=& \int_0^{T_{\rm RH}}\left.\frac{dP(T^{\prime})}{dT^{\prime}}\right|_{\rm eq.\,(\ref{eq:ProbRadiation})} dT^{\prime} \\
&+& \int_{T_{\rm RH}}^{T_{\rm MAX}}\left.\frac{dP(T^{\prime})}{dT^{\prime}}\right|_{\rm eq.\,(\ref{eq:ProbOscillation})} dT^{\prime}~,\nonumber
\end{eqnarray}
and it depends on the reheating temperature and the value of the Hubble parameter at the end of inflation via eq.~(\ref{eq:TMAX}).
Notice that, for a given $T_{\rm RH}$,  the Hubble parameter is characterized by  
the lower bound $H_f^{\rm min}  = [4\pi^3g_*(T_{\rm RH})/45]^{1/2}(T_{\rm RH}^2/M_{\rm P})$; this bound follows from the limit in which the inflaton energy density 
equals the energy density of a thermal bath with temperature $T_{\rm RH}$.
In fig.~\ref{fig:OscillatingPhaseDiagram} we show how the instability bound changes for different values of $T_{\rm RH}$
including the dynamics of reheating.
 For definiteness, we take $H_f = 10^{14}$ GeV. 
As expected, comparing the same values of the reheating temperature analyzed in fig.~\ref{fig:ThermalPhaseDiagramBis}, 
the instability bound 
becomes more stringent including the dynamics of reheating. 
As a benchmark example, the value $T_{\rm RH} = 10^{10}$ GeV---outside the experimental ellipses  in fig.~\ref{fig:ThermalPhaseDiagramBis}---approaches again the edge of the  $3$-$\sigma$ region
if the oscillating phase is included.
\begin{figure}[!htb!]
\centering
 \includegraphics[width = 0.45 \textwidth]{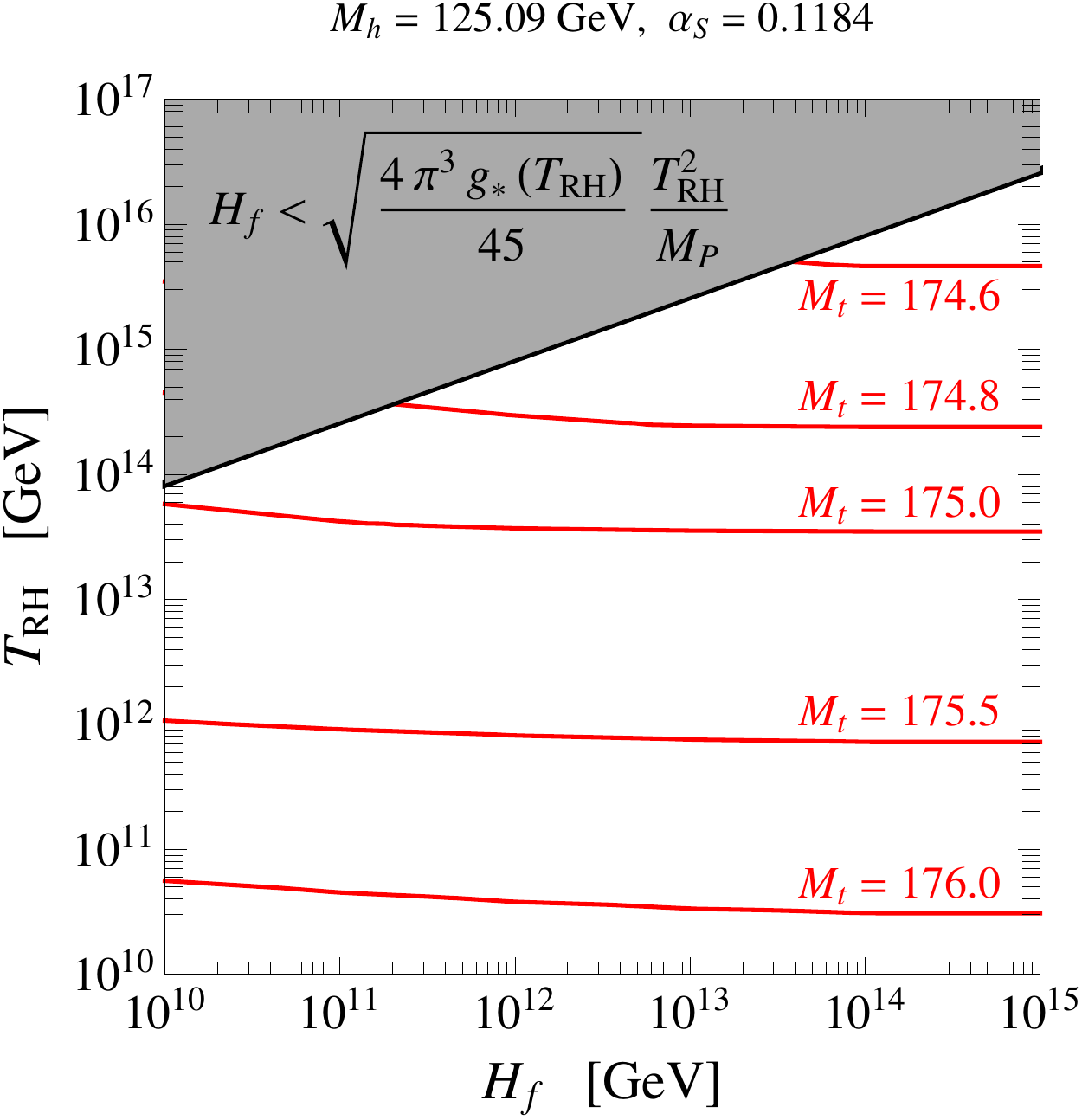}
\caption{ \textit{
Instability bound in the plane $(H_f, T_{\rm RH})$, for different values of the top mass. We include the dynamics of reheating, and we keep fixed $M_h = 125.09$ GeV and $\alpha_s = 0.1184$.
The gray region is excluded by the condition $H_f < H_f^{\rm min}$. For each $M_t$, the region above the corresponding red curve is excluded.
}}
\label{fig:ReheatingBound}
\end{figure}
In order to better investigate the role of the interplay between the reheating temperature and the Hubble parameter at the end of inflation, 
in fig.~\ref{fig:ReheatingBound} we recast the instability bound in the plane $(H_f, T_{\rm RH})$ for different values of the top mass.
For each value of $M_t$,
the values of $T_{\rm RH}$ above the corresponding red curve are excluded.
We notice that the instability bound, for a fixed value of $M_t$, becomes stronger increasing the value of $H_f$; this is expected, since the larger $H_f$ the higher $T_{\rm MAX}$.
However, we also notice that the $H_f$ dependence is very mild (after all $H_f$ enters only as $H_f^{1/4}$ in $T_{\rm MAX}$). 
As for the rest, fig.~\ref{fig:ReheatingBound} retraces what already foreseen in fig.~\ref{fig:OscillatingPhaseDiagram}.
Stringent bounds on the top mass---close to the present experimentally measured central value---can be reached only for very high (yet reasonable) reheating temperatures. 
For reheating temperatures $T_{\rm RH} \simeq 10^{10}$-$10^{11}$ GeV, the bound on the top mass is $M_t \gtrsim 176$ GeV, at the border of the 
experimental $3$-$\sigma$ confidence interval.  

Let us now conclude this section summarizing in a nutshell our results. Thermal corrections 
are relevant for the computation of the instability region in the SM phase diagram, and they can put a very stringent bound on $M_t$ 
close to the present measured central value if also 
the uncertainties on $\alpha_s$ are included.
However, they crucially depend on the temperature of the early Universe.
As already noticed in~\cite{Espinosa:2007qp,Espinosa:2015qea}, therefore, 
the fate of the SM and its cosmological history are inextricably linked.

A crucial question now seems to be: what was the highest temperature  ever recorded in the early Universe after inflation ended?
On a general ground, one could be inclined to think that it must have been very high.
Let us provide one example in the context of thermal leptogenesis and neutrino mass generation via type-I seesaw~\cite{Espinosa:2007qp}. 
On the one hand, in order for baryogengesis to proceed via leptogenesis the mass scale $M$ of the sterile neutrinos must
be of the order of $10^9$ GeV or larger~\cite{Giudice:2003jh,Davidson:2008bu}; on the other one, in order to produce thermally the heavy neutrino states a reheating temperature of the Universe after inflation of $T_{\rm RH} > M$
is required. This simple argument seems to point towards a value of the order of $T_{\rm RH}\gtrsim 10^{10}$ GeV, a temperature high enough
 to generate large thermal corrections, as shown in fig.~\ref{fig:OscillatingPhaseDiagram}.
 
 Moreover, as already stated in the introduction, in~\cite{Espinosa:2015qea} a large reheating temperature after inflation (from $T_{\rm RH} \simeq 10^{7}$ GeV up to $T_{\rm RH} \simeq 10^{17}$ GeV, the actual value
  depending on the instability scale 
 of the Higgs potential and the value of the Hubble constant during inflation)  seems to be suggested by inflation itself, since 
it may tame dangerous quantum fluctuations of the Higgs field.

\section{Conclusions and prospects}\label{sec:Conclusions}

In this paper we revisited and updated 
the computation of thermal corrections to the stability of the electroweak vacuum in the SM.
We followed the approach of~\cite{Espinosa:1995se}, 
based on {\it i)} the computation of the effective potential at finite temperature, and {\it ii)} the exact numerical solution of the bounce equation.
Although the importance of thermal corrections was recently reiterated in~\cite{Espinosa:2007qp,IsidoriTalk,Espinosa:2015qea},
a full computation including the most updated expressions for effective potential, beta functions and matching conditions 
was still missing. Our results can be summarized as follows.

 First, we studied the impact of thermal corrections on the instability of the electroweak vacuum considering the highest allowed cut-off for the temperature, 
$T_{\rm cut-off} \simeq 10^{18}$ GeV. The corresponding SM phase diagram 
is shown in fig.~\ref{fig:ThermalPhaseDiagram} (referred to the parameters $M_h$ and $M_t$),
 and fig.~\ref{fig:ThermalPhaseDiagramAlfa} (referred to the parameters $M_t$ and $\alpha_s$). 
 Thermal corrections turn out to be very important, and they strengthen the constraining power of the instability bound on the SM parameters if compared with the case at $T = 0$.
 If taken at face value, our results show that the instability bound at finite temperature excludes values of the top mass
 $M_t \gtrsim 173.6$ GeV, if $M_h \simeq 125$ GeV, and including the uncertainties on the strong coupling constant at the weak scale. Parametrically, our bound
 is given by eq.~(\ref{eq:BoundMaxT}).

Second, we studied the temperature dependence of the instability bound. Thermal corrections crucially depend on the reheating temperature, hence on the cosmological history of the early Universe after inflation ended.
From this perspective, the case previously studied corresponds to a limit scenario in which $T_{\rm RH}  \simeq 10^{18}$ GeV. 
In order to explore the temperature dependence, 
we investigated two possible situations. 
{\it 1)}  We considered the reheating after inflation as an instantaneous process. 
According to this simplified assumption, the Universe experienced a sharp transition from the inflationary epoch to the radiation-dominated phase.
Our results are shown in fig.~\ref{fig:ThermalPhaseDiagramBis}. 
The instability bound at finite temperature, now cut-offed at $T_{\rm cut-off} = T_{\rm RH}$, weakens. 
However, for $T_{\rm RH} \simeq 10^{11}$ GeV the instability bound still lies at the edge of the $3$-$\sigma$ confidence region for the experimentally measured values of $M_h$ and $M_t$.
For larger values of $T_{\rm RH}$, the SM enters in the instability region.
Parametrically, our bound as a function of $T_{\rm RH}$ is given by eq.~(\ref{eq:BoundTRH}).
{\it 2)} We included in our analysis the dynamics of reheating. The instability bound
becomes stronger if compared with the case of instantaneous reheating
since it includes the oscillating phase of the inflaton field in the interval $T_{\rm RH} \leqslant T \leqslant T_{\rm MAX}$, where $T_{\rm MAX}$ is given by 
eq.~(\ref{eq:TMAX}) and depends on the value of the Hubble parameter at the end of inflation. 
Our results are shown in fig.~\ref{fig:OscillatingPhaseDiagram}. We find that if $T_{\rm RH} \gtrsim 10^{10}$ GeV
the SM starts to fall in the instability region of the phase diagram.

To conclude,
the metastability region of the SM phase diagram considerably shrinks if thermal corrections to the decay of the electroweak vacuum are included. 
On the quantitative level, the impact of these corrections depends on the cosmological history of the early Universe,
 as shown in~\cite{Espinosa:2007qp,Espinosa:2015qea} and discussed in more detail in this paper.
 From a more qualitative perspective, 
unveiling the true nature of  near-criticality becomes an even more urgent question. 
To this end, possible directions include a better measurement of the top quark pole mass---if possible at a future high-energy electron-positron collider---and a deeper 
understanding of the interplay with the physics of the early Universe.

\smallskip

\acknowledgments

We are grateful to Jos\'e Ram\'on Espinosa for many useful advices. We also thank C.~Corian\`o for discussions.
The work of A.U. is supported by the ERC Advanced Grant n$^{\circ}$ $267985$, ``Electroweak Symmetry Breaking, Flavour and Dark Matter: One Solution for Three Mysteries" (DaMeSyFla).

\appendix

\section{Effective potential}\label{app:A}

The effective potential is given by two contributions, the $T=0$ corrections and the thermal effects, computed in the $\overline{\textrm{MS}}$ scheme and in the Landau gauge. For the zero-temperature term we have considered up to the two-loop corrections, but the complete expression is too lengthy to be given here. In order to setup our conventions, we only show the improved tree-level expression and the one-loop terms
%
\begin{equation}
V_0(\phi) = -\frac{1}{4}m^2(t)\phi^2(t) + \frac{1}{4}\lambda(t)\phi^4(t)\approx \frac{1}{4}\lambda(t)\phi^4(t) \label{eq:TreeLevel}
\end{equation}
\begin{equation}
V_{{\rm 1-loop}}(\phi) = \sum_{i = W,Z,t,\chi,h} \frac{n_i}{64\pi^2} m_i(t)^4 \left[
\ln\frac{m_i^2(t)}{\mu^2(t)}  - C_i
\right]~,
\end{equation}
where the coefficients $n_i, \, C_i$ are
\begin{eqnarray}
&& n_W = 6, \,\, n_Z = 3, \,\, n_t = -12, \,\, n_\chi = 3, \,\, n_h = 1, \nonumber \\
&& C_W = C_Z = 5/6, \,\, C_t = C_\chi = C_h = 3/2,
\end{eqnarray}
while the mass parameters are given by 
\begin{eqnarray}
m_W^2(t) &=& \frac{1}{4}g^2(t)\phi^2(t)~,\\   
m_Z^2(t) &=& \frac{1}{4}\left[g^2(t) + g^{\prime 2}(t)\right]\phi^2(t)~,   \\  
m_t^2(t) & =& \frac{1}{2}y_t^2(t)\phi^2(t)~,\\
m_{\chi}^2(t) &=& -\frac{m^2(t)}{2} + \lambda(t)\phi^2(t) \approx \lambda(t)\phi^2(t)~,\\
m_{h}^2(t) &=& -\frac{m^2(t)}{2} + 3\lambda(t)\phi^2(t) \approx 3\lambda(t)\phi^2(t)~.
\end{eqnarray}
Since we are interested in large field values, we neglect the quadratic term in the Higgs potential. 
All the SM parameters are running with the three-loop RG equations, so that our analysis takes into account all the NNLL contributions. \\
The one-loop thermal corrections to the effective potential are (see~\cite{Quiros:1994dr} for a thorough discussion)
\begin{widetext}
\begin{eqnarray}
V_{{\rm 1-loop}}(\phi, T)&=&
\sum_{i = W,Z,\chi,h} \frac{n_iT^4}{2\pi^2} J_{\rm B}\left(
\frac{m_i^2(t)}{T^2}
\right) +
\frac{n_t T^4}{2\pi^2} J_{\rm F}\left(
\frac{m_t^2(t)}{T^2}
\right)~, \\
V_{{\rm ring}}(\phi, T) &=& \sum_{i = W_L,Z_L,\gamma_L, \chi, h}
\frac{n_i T^4}{12\pi}\left\{
\left[\frac{m_i^2(t)}{T^2}\right]^{3/2}
-
\left[\frac{\mathcal{M}_i^2(\phi)}{T^2}\right]^{3/2}
\right\}~.\label{eq:Rings}
\end{eqnarray}
\end{widetext}
The thermal integrals are 
\begin{eqnarray}
J_B(y) &=& \int_0^{\infty}
dx\,x^2\,\ln\left[
1- e^{-\sqrt{x^2 + y}}
\right]~,\label{eq:JB}\\
J_F(y) &=& \int_0^{\infty}
dx\,x^2\,\ln\left[
1+ e^{-\sqrt{x^2 + y}}
\right]~.\label{eq:JF}
\end{eqnarray}
The one-loop thermal potential is improved by the one-loop ring resummation of daisy diagrams in which only the bosonic degrees of freedom are taken into account and, in particular, only the longitudinal component of the vector fields. The degeneracy coefficients are
\begin{equation}
n_{W_L} = 2, \qquad n_{Z_L} = 1, \qquad n_{\gamma_L} = 1~.
\end{equation} 
The Debye masses are $\mathcal{M}_i^2(\phi) = m_i^2(t) + \Pi_i(\phi,T)$, with the following temperature-dependent self-energies
\begin{eqnarray}
\Pi_h(\phi,T)&=& \left(
\frac{3g^2 + g^{\prime 2}}{16} +\frac{\lambda}{2} + \frac{y_t^2}{4}
\right)T^2 = \Pi_{\chi}(\phi, T)~,\nonumber \\
\Pi_{W_L}(\phi,T)&=& \frac{11}{6}g^2T^2~,\nonumber\\
\Pi_{W_T}(\phi, T) &=& \Pi_{Z_T}(\phi, T)  = \Pi_{\gamma_T}(\phi, T) = 0~, 
\end{eqnarray}
where we omit the $t$-dependence implied by the RG improvement. Finally, mapping $(W_3, B)$ into $(Z, \gamma)$, we find
\begin{eqnarray}
\mathcal{M}_{Z_L}^2(\phi) &=& \frac{1}{2}
\left[
m_Z^2(t) + \frac{11}{6}\frac{g^2}{\cos^2\theta_W}T^2 + \Delta(\phi, T)
\right]~,\nonumber\\
\mathcal{M}_{\gamma_L}^2(\phi) &=& \frac{1}{2}
\left[
m_Z^2(t) + \frac{11}{6}\frac{g^2}{\cos^2\theta_W}T^2 - \Delta(\phi, T)
\right]~,\nonumber \\
\end{eqnarray}
with
\begin{eqnarray}
\Delta^2(\phi,T) &=& m_Z^4(t) + \frac{11}{3}\frac{g^2\cos^2 2\theta_W}{\cos^2\theta_W}\times \nonumber \\
&& \left[
m_Z^2(t) + \frac{11}{12}\frac{g^2}{\cos^2\theta_W}T^2
\right]T^2~.
\end{eqnarray}
Having set the formalism, let us now quantify the impact of the ring corrections in eq.~(\ref{eq:Rings}).
\begin{figure}[!htb!]
\centering
 \includegraphics[width = 0.45 \textwidth]{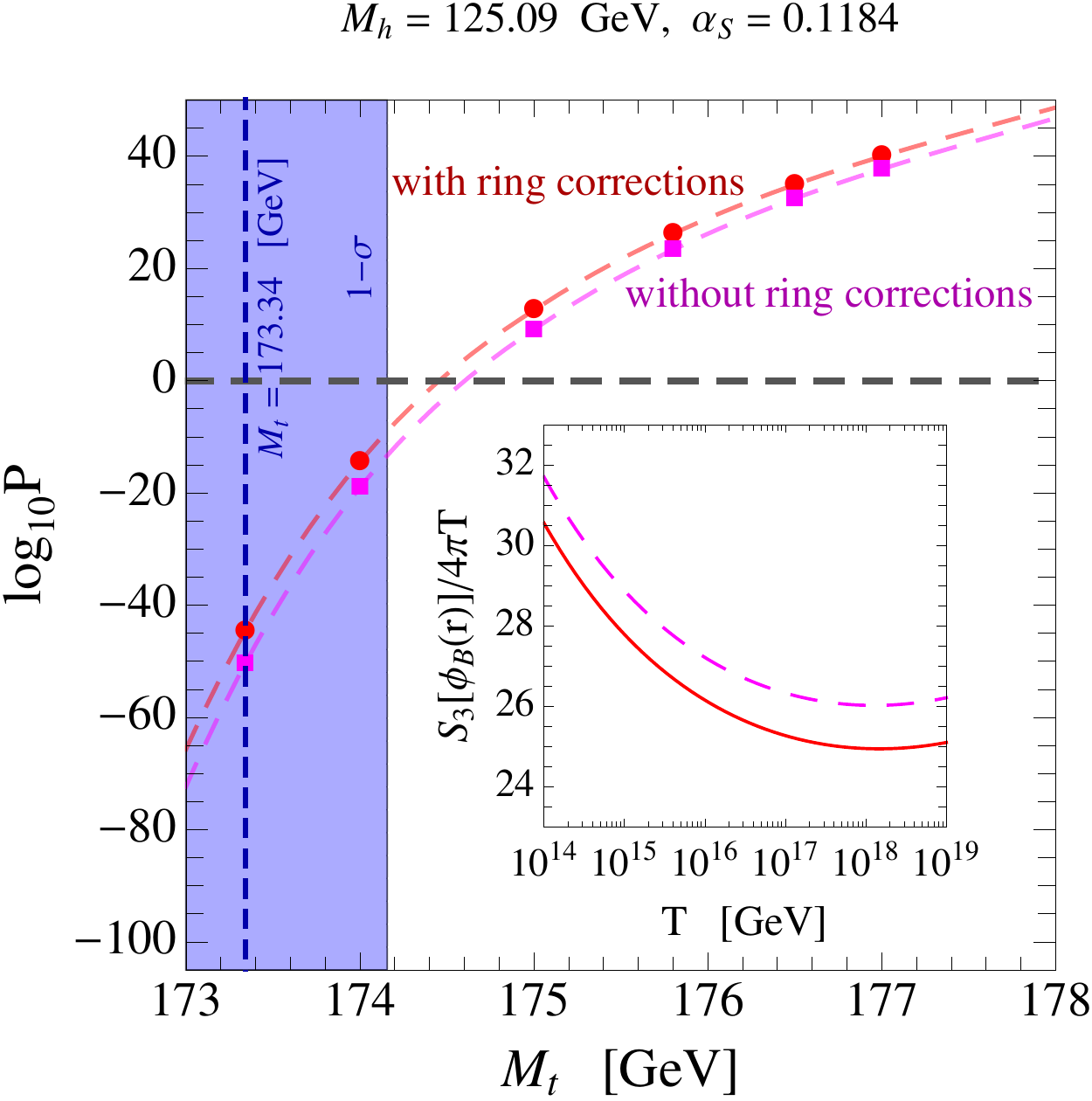}
\caption{ \textit{
Same as in fig.~\ref{fig:ThermalPhaseDiagram} but considering only $M_h = 125.09$ GeV, and comparing the case  with (red, filled circles) and without (magenta, filled squares) ring corrections in the effective potential. In the insert, we compare the bounce action (same as in fig.~\ref{fig:BounceEnergy}, left panel) with (solid, red) and without (dashed, magenta) ring corrections.
}}
\label{fig:Ring}
\end{figure}
These corrections take into account the resummation  of daisy diagrams. 
As clear from eq.~(\ref{eq:Rings}), the ring contribution vanishes in the limit $\phi \gg T$.
The numerical approach carried out in section~\ref{sec:Bounce} 
showed that, for a given $T$, the thermal tunneling always occurs at field value $\phi_B(0)\sim 10\times T$. 
This is enough to argue that ring contributions do not play a crucial role.
Fig.~\ref{fig:Ring}---where we compute the total probability in eq.~(\ref{eq:ThermalProb}) with (red) and without (magenta) ring contributions---confirms this hypothesis.
Ring contribution generate a $\sim 5\%$ correction to the bounce action (insert plot in fig.~\ref{fig:Ring}); 
in turn, this correction translates into a $\sim 0.2$ GeV strengthening of the instability bound.

\begin{figure}
\centering
 \includegraphics[width = 0.45 \textwidth]{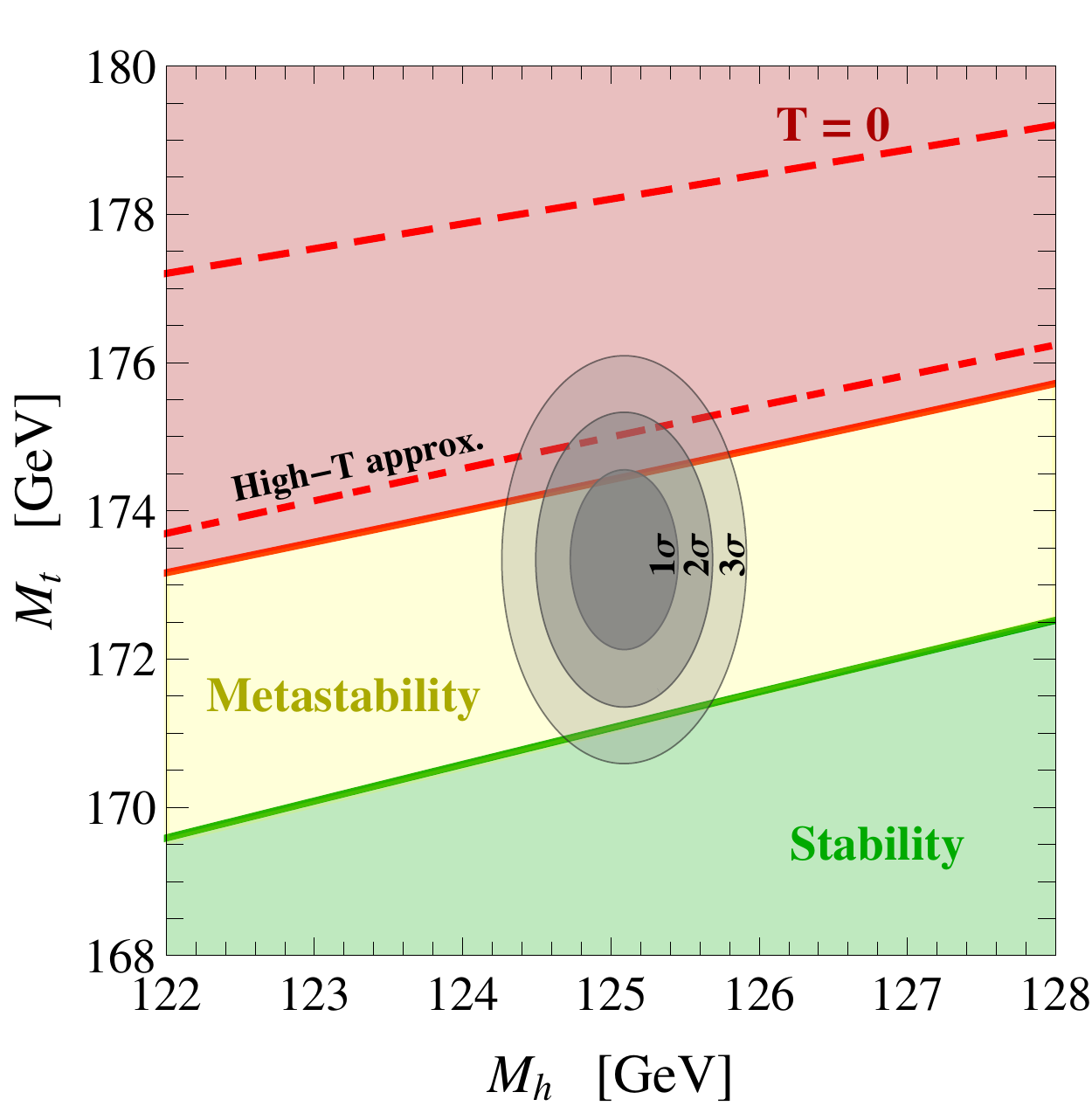}
\caption{ \textit{Same as fig.~\ref{fig:ThermalPhaseDiagram}. The dot-dashed line corresponds to the instability bound in the high-temperature approximation.
}}
\label{fig:ThermalPhaseDiagramVokos}
\end{figure}

The analysis presented in this work is performed by numerical methods and, as such, does not rely on any analytical approximation. It is interesting, therefore, to compare our numerical results to those obtained, for instance, in the large $T$ regime~\cite{Arnold:1991cv}. In the high-temperature limit the effective potential can be written in the form
\begin{equation}
V_{\rm eff}(\phi,T) \simeq \frac{\lambda_{\rm eff}}{4} \phi^4 + \frac{1}{2} \kappa^2 \phi^2 T^2 + {\rm const}
\end{equation}   
where the constant term, being $\phi$-independent, can be neglected and the coefficient $\kappa^2$ is
\begin{eqnarray}
\kappa^2 &=& \frac{1}{12} \left( \frac{3}{4} g'^2 + \frac{9}{4} g^2 + 3 y_t^2 + 6 \lambda \right) \nonumber \\
&-& \frac{1}{32 \pi}  \sqrt{\frac{11}{6} }\left( g'^3 + 3 g^3 \right) \nonumber \\
&-& \frac{3}{16\pi} \lambda \sqrt{g'^2 + 3 g^2 + 8 \lambda + 4 y_t^2}~.
\end{eqnarray}
The first line comes from the high-$T$ expansion of the thermal integrals while the last two from the ring potential. For large field values, the effective potential at $T=0$ can be expressed in terms of an effective quartic coupling $\lambda_{\rm eff}$ which accounts for one- and two-loop corrections. With such a simple expression for the effective potential, the bounce equation can be solved straightforwardly, obtaining~\cite{Arnold:1991cv} $S_3[\phi_B(r)] \simeq  - (6.015) \pi \kappa/\lambda_{\rm eff} T$.
We recall once again that all the parameters are scale dependent and run with the RG equations. To minimize the impact of large logs, the renormalization scale is chosen to be equal to the canonical normalized scalar field. Moreover, as we have already shown, the thermal tunneling is characterized by a field value roughly of the order of the temperature,  
thus reducing the analysis to a problem with just one scale, fixed by the temperature. We show in fig.~\ref{fig:ThermalPhaseDiagramVokos} the SM phase diagram within this high-temperature approximation. This results in a less tight instability bound of $\sim 0.6$ GeV with respect to the full numerical analysis. 

\section{Beyond leading order thermal corrections}\label{app:B}

In this paper we truncated the perturbative expansion of the effective potential at finite temperature at one-loop (including resummed ring diagrams). 
In this appendix we discuss, at the qualitative level, the validity of this description together with possible future improvements. 
The effective potential at finite temperature enters in the euclidean action in eq.~(\ref{eq:SE}).
Since at zero temperature we worked at two-loop order, it is 
natural to ask what is the impact of two-loop thermal corrections. 
At two loops, thermal corrections to the effective potential were studied in~\cite{Bagnasco:1992tx,Arnold:1992rz,Fodor:1994bs} in the context of
the electroweak phase transition.
Here what we want to stress is that adding two-loop thermal corrections to the effective potential at finite temperature 
does not improve the precision 
of the computation, since the one-loop result is already plagued by theoretical uncertainties---very likely of the same order of the two-loop 
corrections.
The crucial point is that the euclidean action in eq.~(\ref{eq:SE}) relies on different approximations. 
Before proceeding, we stress that 
a comprehensive  analysis of the theoretical errors associated with the computation
of the stability of the electroweak vacuum at finite temperature is an extremely difficult task---well beyond the purposes of this paper and, to the best of
our knowledge, never studied before in the literature.
In what follows, we highlight 
the most relevant aspects of such analysis. 

\begin{figure*}[!htb!]
\centering
  \begin{minipage}{0.45\textwidth}
   \centering
   \includegraphics[scale=0.6]{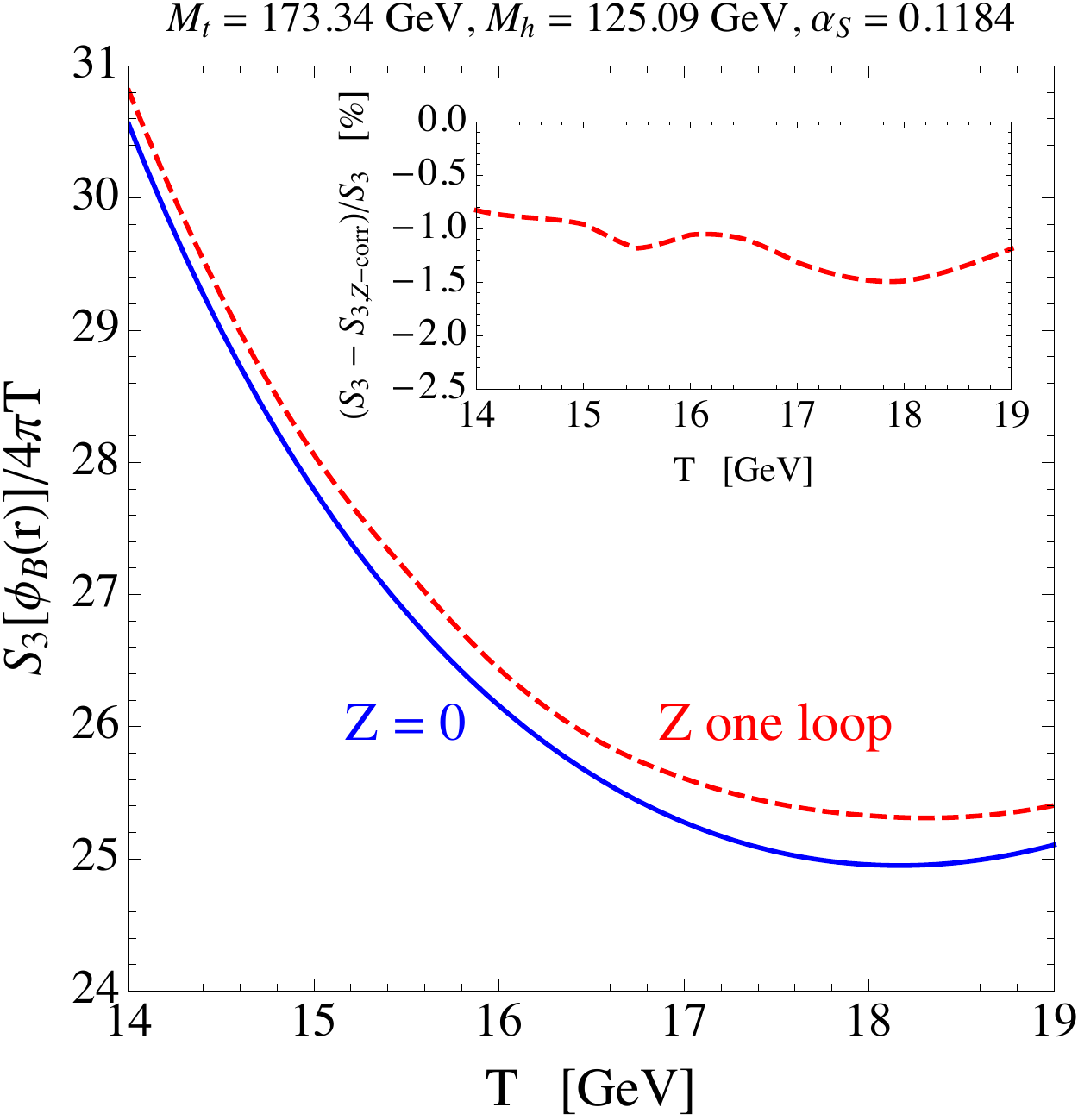}
    \end{minipage}\hspace{0.3 cm}
   \begin{minipage}{0.45\textwidth}
    \centering
    \includegraphics[scale=0.6]{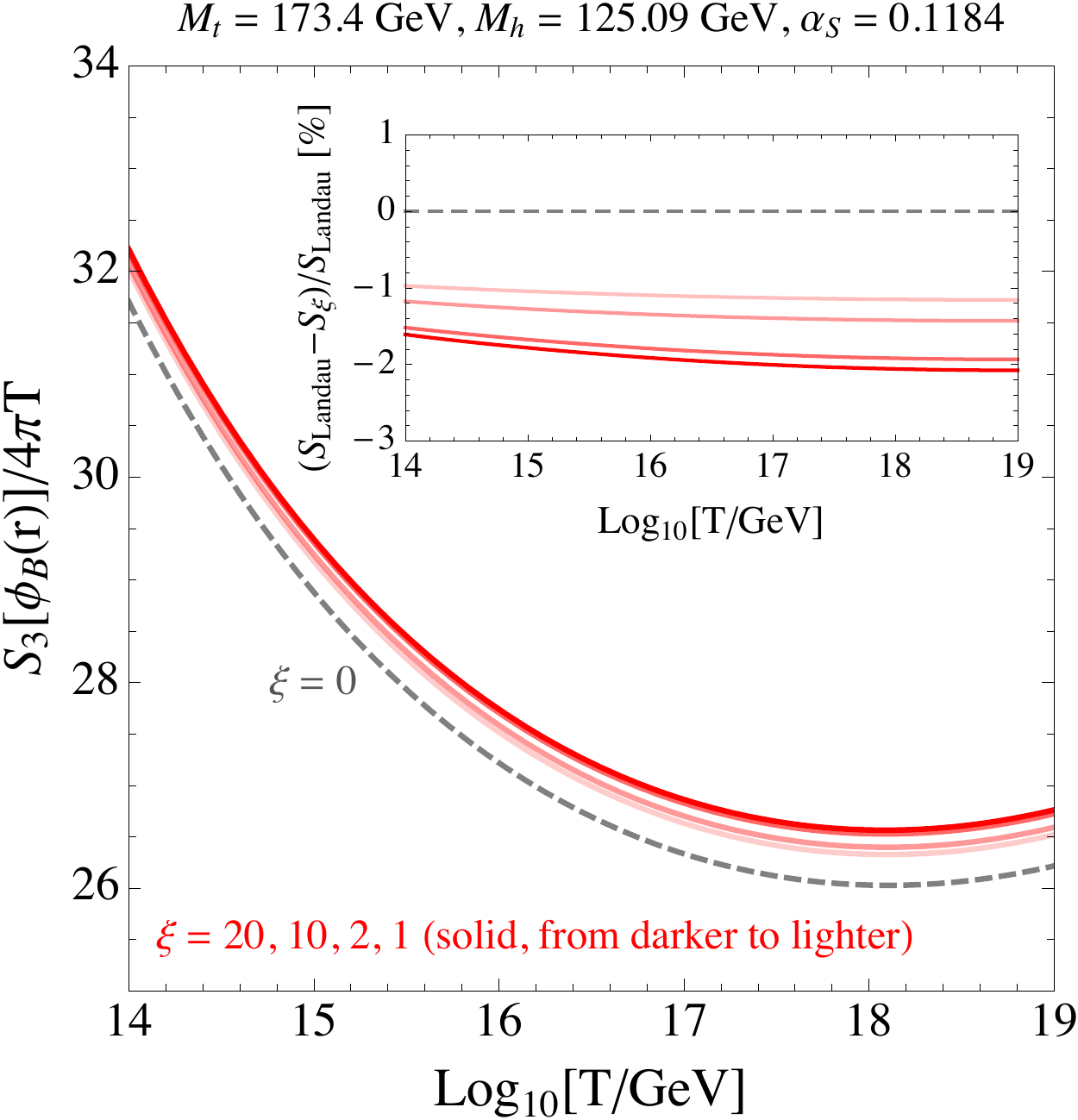}
    \end{minipage}
\caption{\small \textit{
Euclidean action of the bounce solution as a function of the temperature.
Left panel: we compare the case without (blue solid line) and with (red dashed line) the one-loop correction to the kinetic term, eq.~(\ref{eq:ActionImproved}).
Right panel: to test gauge-dependence, we compare the impact of different choices of $\xi$ for the one-loop thermal corrections to the effective potential.
}}\label{fig:Z2Test}
\end{figure*}

\begin{itemize}

\item[$\circ$] {\it High-temperature approximation}. At sufficient large temperature, in the computation 
of the euclidean action the integration over the euclidean time amounts to multiply by $T^{-1}$ the three dimensional action 
corresponding to the $O(3)$ symmetric bubble~\cite{Linde:1980tt}
\begin{equation}\label{eq:Approx}
S_{\rm E}[\phi_B(r)] = T^{-1}S_3[\phi_B(r)]~,
\end{equation}
with $S_3[\phi_B(r)]$ as in eq.~(\ref{eq:SE}). The parameter controlling this approximation is 
the inverse of the bounce size at zero temperature~\cite{Linde:1980tt}.
In section~\ref{sec:Bounce} we computed this quantity, and we found $R_{M}^{-1} \simeq 2.8\times 10^{16}$ GeV. 
From the right panel in fig.~\ref{fig:BounceEnergy}, 
we see that the decay probability is dominated by larger values of temperature. 
However, the validity of the approximation in eq~(\ref{eq:Approx}) is not always guaranteed and deserves further studies. 

\item[$\circ$] {\it Corrections to the kinetic term}. In the  computation 
of the euclidean action one should also include---in addition to the one-loop thermal corrections to the effective potential---one-loop corrections
to the kinetic term. In full generality, 
these corrections  can be written in a gradient expansion in power of derivative of the classical background field $\phi$~\cite{Bodeker:1993kj}
\begin{align}\label{eq:ActionImproved}
S_{\rm E}[\phi(r,t)] &= \int_{\beta} 
\left[
\frac{1}{2}\left(\partial_{\mu}\phi\right)\left(\partial^{\mu}\phi\right) + V_{\rm eff}(\phi, T)
\right] \nonumber \\ &+ 
\int_{\beta}\sum_{n = 2}^{\infty}
\frac{1}{n!}Z_{n}(\phi, T)\left(\partial_{\mu}\phi\right)^n~,
\end{align}
where each $Z_{n}(\phi, T)$, in turn, can be expanded in a power series in the couplings. The euclidean space-time integration is
\begin{equation}
\int_{\beta} \equiv \int_0^{\beta}dt \int d^3 \vec{x}~,~~~~\beta \equiv 1/T~,
\end{equation}

\item[$\circ$] {\it Gauge dependence}. If the corrections to the kinetic term are neglected, 
the effective action becomes gauge-dependent as a consequence of a broken Nielsen identity~\cite{Garny:2012cg}. 

\end{itemize}

In fig.~\ref{fig:Z2Test} we estimated the impact of gauge dependence and corrections to the kinetic term.
In the left pane, we included the corrections to the kinetic term 
truncating at first order the gradient expansion in eq.~(\ref{eq:ActionImproved}), 
and computing at one-loop the wave-function renormalization $Z_2(\phi,T)$ following~\cite{Bodeker:1993kj}.
In the right panel, we estimated the impact of gauge dependence at finite temperature at one-loop in a generic $R_{\xi}$ gauge.
In both cases we found a correction to the effective action of the bounce of the order of few {\it percent}.

A more detailed  analysis of these corrections will be presented in a forthcoming work.

\end{document}